\documentclass[a4paper,fleqn,usenatbib]{mnras}

\usepackage{newtxtext,newtxmath}

\usepackage[T1]{fontenc}
\usepackage{ae,aecompl}

\usepackage{subfig,graphicx}	
\usepackage{amsmath}	
\usepackage{amssymb}	
\usepackage{upgreek}
\usepackage{pdflscape}
\title[First results from the LIFE project]
{First results from the LIFE project: discovery of two magnetic hot evolved stars\thanks{Based on observations obtained at the Canada-France-Hawaii 
Telescope (CFHT) operated by the National Research Council of Canada,  
the Institut National des Sciences de l'Univers of the CNRS of France, 
and the University of Hawaii.}}

\author[A.~J. Martin]{A.~J. Martin$^{1}$\thanks{Contact e-mail: \href{mailto:alexander.martin@obspm.fr}
{alexander.martin@obspm.fr}}, 
C. Neiner$^{1}$, 
M.~E. Oksala$^{1,2}$, 
G.~A. Wade$^{3}$, 
Z. Keszthelyi$^{3,4}$,
\newauthor{L. Fossati$^{5}$,
W. Marcolino$^{6}$, 
S. Mathis$^{7,1}$,
C. Georgy$^{8}$}\\
$^{1}$LESIA, Observatoire de Paris, PSL Research University, CNRS, Sorbonne Universit\'{e}s, UPMC Univ. Paris 06, \\
Univ. Paris Diderot, Sorbonne Paris Cit\'{e}, 5 place Jules Janssen, F-92195 Meudon, France\\
$^{2}$Department of Physics, California Lutheran University, 60 West Olsen Road 3700, Thousand Oaks, CA, 91360, USA\\
$^{3}$Department of Physics, Royal Military College of Canada, PO Box 17000, Station Forces, Kingston, Ontario, K7K 7B4, Canada\\
$^{4}$Department of Physics, Engineering Physics and Astronomy, Queen's University, 99 University Avenue, Kingston, ON K7L 3N6, Canada\\
$^{5}$Space Research Institute, Austrian Academy of Sciences, Schmiedlstrasse 6, A-8042 Graz, Austria\\
$^{6}$Universidade Federal do Rio de Janeiro, Observat\'{o}rio do Valongo, Ladeira Pedro Antonio, 43, CEP 20.080-090, Rio de Janeiro, Brazil\\
$^{7}$Laboratoire AIM Paris-Saclay, CEA/DRF - CNRS - Universit\'{e} Paris Diderot, IRFU/DAp Centre de Saclay, 91191 Gif-sur-Yvette,
France\\
$^{8}$Geneva Observatory, University of Geneva, chemin des Maillettes 51, 1290 Sauverny, Switzerland}

\date{Accepted XXX. Received YYY; in original form ZZZ}

\pubyear{2017}

\begin{document}
\label{firstpage}
\pagerange{\pageref{firstpage}--\pageref{lastpage}}
\maketitle

\begin{abstract}
We present the initial results of  the Large Impact of magnetic Fields on the 
Evolution of hot stars (LIFE) project. The focus of this project is the search for 
magnetic fields in evolved OBA giants and supergiants with 
visual magnitudes between 4 and 8, with the aim to investigate how 
the magnetic fields observed in upper main sequence (MS) stars evolve from the MS until the late post-MS stages. In this paper, we  present spectropolarimetric 
observations of 15 stars observed using the ESPaDOnS instrument of the CFHT. 
For each star, we have determined the fundamental parameters and  
have used stellar evolution models to calculate their mass, age and radius.
Using the LSD technique, we have produced averaged line profiles for each star. From these profiles, 
we have measured
the longitudinal magnetic field strength and have calculated the detection probability. 
We report the detection of magnetic fields in two stars of our sample: 
a weak field of $B_l=1.0\,\pm\,0.2$\,G is detected in the post-MS A5 star 19\,Aur and a stronger field of 
$B_l=-230\,\pm\,10$\,G is detected in the MS/post-MS B8/9 star HR\,3042. \\
\end{abstract}



\section{Introduction}

The last decade has seen a remarkable increase in our knowledge of the properties of magnetic 
fields in main sequence (MS) OB type stars. As a result of a number of surveys 
\citep[e.g.,][]{Fossati2015,Fossati2015b,Wade2016,Grunhut2017},  
convincing evidence indicates that $\sim$10\% of MS OB stars are magnetic. 
This is consistent with the findings  
for MS A stars \citep{Power2008,Auriere2007}. MS OB stars 
host roughly dipolar, often oblique fossil magnetic fields 
\citep{Neiner2015,Grunhut2015}. There is currently no known mechanism to generate and sustain an efficient  
dynamo field in a radiative envelope, which would affect the global fossil field during the MS phase of 
OB type stars \citep{Zahn2007,Rudiger2012, Neiner2015}. The observed fossil fields  likely formed in the pre-MS phases: 
during its formation the star could ensnare the weak field present in the interstellar medium 
(ISM) and dynamo fields generated during pre-MS stages can enhance this seed field \citep{Mestel1999,Alecian2008}. 
The field then relaxes to a stable configuration, which is mainly dipolar at the surface \citep{Braithwaite2004,Duez2010}.

Since no dynamo field is observed in MS OB stars, 
there is nothing to replenish the field over time. It has been shown by 
 \citet{Bagnulo2006,Landstreet2007,Landstreet2008} and \citet{Fossati2016}, that the 
magnetic fields at the surface of stars decrease with stellar age on the MS. 
This decrease in strength likely occurs in part as a result of flux conservation in response to the 
increase in stellar radius, possibly altered by Ohmic decay or other currently unknown effects.

Furthermore, it is likely that the stellar magnetic field has 
 a strong influence on the evolution of the host star \citep{Moss1984,Langer2012,Maeder2014,Keszthelyi2017,Georgy2017}. 
The effects of the field can be separated into those operating at the surface of the star and 
those in the interior. Theoretical models predict that interactions between the magnetic field 
and the stellar wind can reduce the surface mass-loss and stellar rotation rate \citep{udDoula2002,udDoula2008,udDoula2009,Meynet2011,Bard2016}.  
Indeed, studies of massive stars show that the stellar wind can become  
trapped in a magnetosphere rigidly coupled to the star's magnetic field \citep{Landstreet1978,Oksala2015}. Conservation of 
angular momentum then leads to a decrease in the rotation rate of the star \citep[e.g.,][]{Townsend2010,Mikulavsek2007}. In the stellar interior, the magnetic field has the potential to affect the mixing and diffusion of chemical elements,  
the internal rotational profile and angular momentum 
\citep{Mestel1999,Mathis2005,Briquet2012,Sundqvist2013,Maeder2014,Stift2016}.

In general, however, magnetic fields have not been taken into account in evolutionary models, except 
for Taylor-Spruit dynamo fields \citep[e.g.,][]{Maeder2003,Heger2005}, which are themselves  inconsistent 
with those observed in 
MS OBA type stars. In particular, the observed fields are not correlated with rotation as would be expected for an $\upalpha-\upOmega$ dynamo like that proposed by \citet{Spruit2002}. 
It is therefore important to study these stars in detail to provide observational constraints to stellar evolution models. However, the structural changes which occur in a star during the MS are, in general, not sufficient to allow us to understand how  
magnetic fields change as stars evolve and how stars respond to the presence of a 
magnetic field. 

One potential pathway towards exploring the evolution of magnetic fields in 
massive stars lies in the study of evolved OBA stars.  We already have convincing 
evidence that $\sim$10\% of MS OB stars have magnetic fields 
\citep{Grunhut2015,Fossati2015,Grunhut2017}, and the same appears to be true for PMS stars \citep{Alecian2013}.  It is reasonable to suspect that the 
same may well be true for evolved descendants. 
These stars not only 
 provide the possibility to study the evolution of magnetic fields over a longer evolutionary
time-frame, but also to investigate the evolution of the magnetic field in response to the 
significant changes in stellar structure occurring during the star's transition through the post-MS. 

As the star moves through the post-MS, its radius vastly increases \citep[e.g.,][]{Ekstrom2012,Georgy2013,Sanyal2015}. 
Initially, the fossil field structure should remain unchanged, since the radiative envelope still exists and 
there is  no mechanism to allow for the formation of a dynamo. 
 However, as the 
star continues to evolve, convective zones form in the upper regions of the 
stellar envelope \citep[e.g.,][]{Neiner2017}. These regions have the potential to generate dynamo magnetic fields. These 
new dynamo fields might have a significant impact on the fossil field already 
present in the star \citep{Featherstone2009,Auriere2008}. Indeed the studies of FGK type giants and supergiants show magnetic fields powered 
by dynamos \citep[e.g.,][]{Grunhut2010,Auriere2015}. These stars are the evolutionary descendants of
OBA type MS stars and so it is possible that a star originally with a fossil magnetic field on 
the MS experiences a period with both a fossil field and dynamo field, and finally evolves 
to a state in which only the dynamo field signature can be observed. In fact, the intermediate mass red giant star EK Eri shows tantalising evidence that it hosts both a dynamo field and the remnant of an Ap fossil field \citep{Auriere2011}.

Until very recently, no post-MS magnetic OBA stars had been unambiguously identified, despite high-precision studies having been conducted \citep[e.g.,][]{Verdugo2005,Shultz2014}. The O9.5 supergiant $\upzeta$\,Ori\,A was found to have a magnetic field \citep{Bouret2008,Blazere2015}, 
however, this star was shown to likely be still on the MS \citep{Fossati2015b}, despite its supergiant classification. A magnetic field was also 
detected in the B1.5 star $\upepsilon$\,CMa by \citet{Fossati2015b,Neiner2017} showed however, that this star is also located at the end of its MS phase.
As part of the BRIght Target Explorer spectropolarimetric survey  
\citep[BRITEpol;][]{Neiner2016}, 
\cite{Neiner2017} identified two magnetic A7 supergiants: $\upiota$\,Car and HR\,3890. Luminosity 
measurements and the fundamental parameters associated with these stars 
indicate that they are well into the post-MS phase of evolution. The detected longitudinal magnetic field strengths were found to be roughly 10 and 1\,G for $\upiota$\,Car and HR\,3890, respectively. As a result, we infer that any survey would need to be able to detect 
magnetic fields to a precision of better than 1\,G, in order to determine the incidence rate of magnetic fields in evolved OBA type stars.

To this end, 
we have started the Large Impact of magnetic Fields on the 
Evolution of hot stars (LIFE) project. We are observing the circularly polarised spectrum and measuring the longitudinal 
magnetic field strength ($B_{\rm l}$) of OBA type stars between 
V\,=\,4\,mag and V\,=\,8\,mag with luminosity classes I-III. 
In this paper, we present the analysis of the first 15 stars observed for this study.  
Section~\ref{sec:LIFE} gives details of the LIFE project, the observations and data 
reduction. The fundamental parameters we infer for each star are given in 
Section~\ref{sec:stelparams} along with the evolutionary status of each star. 
In Section~\ref{sec:mag} we give details of the magnetic field detections  
and finally in Sections~\ref{sec:dis} and \ref{sec:conc} we discuss our findings 
and conclude the paper.

\section{The LIFE project}
\label{sec:LIFE}
\begin{table*}
	\caption{Observation log for the initial LIFE targets observed using ESPaDOnS. 
The name of the star, its Henry Draper catalogue (HD) designation and its Johnson V magnitude \citep{Perryman1997} is given. The date of observation and the sequence 
of exposures is shown, where the first value is the number of consecutive polarimetric sequences, the second is the 
four observations taken with different rotations of the Fresnel rhombs and the last is the number of seconds 
per exposure. The next column is the
Heliocentric Julian Date at the middle of the observation (mid-HJD - 2450000). The average S/N of the spectropolarimetric 
sequence per 1.8\,kms$^{-1}$ pixel at $\sim$5000\,\AA\ is shown, along with the mean S/N of the LSD $I$ and $V$ profiles computed as the mean of the square root of the diagonal elements of the inverse of the autocorrelation matrix \citep[see][]{Donati1997}, scaled by the rms of the fit between the LSD and observed spectra. Finally, the 
number of lines selected for the LSD line mask following the cleaning described in Section~\ref{sec:LSD}.}        
	\label{Table:LIFE_Targets} 
	\centering                               
\begin{tabular}{llccrlcccc}       
		\hline\hline
		Star		& HD No. & Visual & \multicolumn{1}{c}{Date	}	& \multicolumn{1}{c}{Exp. Seq.}	&	mid-HJD	&	Average & LSD $I$ & LSD $V$ & Lines\\
				&		&	Magnitude & (UT)			& \multicolumn{1}{c}{(s)}			& -2450000&S/N&S/N&S/N& in LSD\\   
		\hline                      
		             13\,Mon	&      46300 	&       4.47 & Feb 18, 2016 &    2x4x138&  7437.8113&        758&       3536&      13603&       1394\\
		             15\,Sgr	&     167264 	&       5.29 & May 14, 2016 &    2x4x344&  7524.1189&        911&       1610&       6816&         69\\
			& 	&& May 17, 2016 &    4x4x344&  7526.9869&        863&       2229&       9234&         69\\
			& 	&& May 18, 2016 &    2x4x344&  7528.0034&        921&       1552&       6885&         69\\
		             19\,Aur	&      34578 	&       5.05 & Sep 18, 2016 &    5x4x254&  7651.0154&       1034&       4031&      55620&       4298\\
			& 	&& Oct 20, 2016 &    5x4x254&  7683.0600&       1067&       3808&      56872&       4297\\
		             24\,CMa	&      53138 	&       3.02 & Feb 28, 2016 &    23x4x40&  7447.8037&        983&       4466&      48600&        538\\
		       $\upeta$\,Leo	&      87737 	&       3.48 & Feb 21, 2016 &     5x4x50&  7440.8394&        535&       3735&      17673&       1781\\
		     $\upgamma$\,CMa	&      53244 	&       4.11 & Dec 20, 2016 &   12x4x143&  7744.0163&       1111&      14049&      60259&       1293\\
		           HD\,10362	&      10362 	&       6.33 & Sep 17, 2016 &    1x4x991&  7649.9784&       1187&       2270&      13094&        476\\
			& 	&& Sep 19, 2016 &    1x4x991&  7651.8850&       1238&       2326&      13619&        476\\
		           HD\,42035	&      42035 	&       6.55 & Sep 21, 2016 &   1x4x1313&  7654.0594&       1316&       4327&      22097&       2649\\
		          HD\,186660	&     186660 	&       6.47 & Oct 13, 2016 &   1x4x1122&  7675.7705&       1112&       1850&       7975&        327\\
		          HD\,188209	&     188209 	&        5.6 & Jun 19, 2016 &    8x4x443&  7560.0473&       1017&       1669&      19678&        227\\
		          HD\,209419	&     209419 	&       5.79 & Oct 12, 2016 &    1x4x666&  7674.8808&       1035&       4305&      13777&        600\\
		          HIP\,38584	&      64827 	&       6.85 & Mar 13, 2017 &   1x4x1049&  7826.8619&        728&       4861&      13323&        864\\
		            HR\,3042	&      63655 	&       6.23 & Dec 14, 2016 &    3x4x981&  7738.0787&       1233&       5628&      21271&        201\\
		             PT\,Pup	&      61068 	&       5.69 & Dec 20, 2016 &    1x4x559&  7743.9491&       1105&       1330&      13114&        304\\
		           V399\,Lac	&     210221 	&       6.17 & Jun 14, 2016 &    4x4x604&  7555.0612&        846&       2447&      29738&       1631\\
		\hline          	                                  
	\end{tabular}
\end{table*}

Through the LIFE project, we aim to measure the distribution of  
magnetic field strengths in evolved stars and compare our findings with the results for MS OB stars.
Any detected field will be fully characterised and the 
results will be compared with those of MS stars to determine how the magnetic 
fields of stars change as stars evolve. From this project we aim to provide important 
observational constraints on theories of how magnetic fields and stellar evolution affect and interact with each other.

Choosing the appropriate exposure time for each star is critically important to give us the 
best possible chance of detecting magnetic fields for our sample of stars. 
As a 
result of the MiMeS \citep{Wade2016}, BinaMIcS \citep{Alecian2015} and 
BRITEpol \citep{Neiner2016} surveys we have an accurate exposure time
relationship \citep{Wade2016} which  predicts the spectral signal-to-noise ratio (S/N) 
required to detect magnetic fields to a certain precision, given the spectral type and 
rotational velocity of a star\footnote{To plan the observations, we considered the rotational velocities found through Vizier. If none were present for a given star, we assumed a value of 50\,km\,s$^{-1}$.}. It is then straightforward  
to obtain the necessary exposure time required for each star. 

However, we cannot know a priori the current dipole magnetic field strength ($B_{\textrm{d,current}}$) of the stars; therefore it is necessary to adopt a plausible distribution for the magnetic field. 
We calculate this by comparing the current radius of the star ($R_{\textrm{current}}$)
with its radius at the zero-age main sequence ($R_{\rm ZAMS}$) 
assuming that the magnetic flux is conserved. 
We calculate the expected current surface magnetic field strength to be
\begin{equation}
B_{\textrm{d,current}}= B_{{\rm d,ZAMS}}\left(\frac{R_{\rm ZAMS}}{R_{\textrm{current}}}\right)^2,
\end{equation}
where $B_{{\rm d,ZAMS}}$ is the magnetic field strength at the ZAMS. 
The choice of $B_{{\rm d,ZAMS}}$ is important, because we are 
limited by total exposure time. For this work, we assume that $B_{{\rm d}({\rm ZAMS})} = 0.63$\,kG 
or if this would lead to an exposure time longer than one night, then $B_{{\rm d}({\rm ZAMS})} = 1.4$\,kG. 
These values correspond to 95\% and 85\% completeness of the observed distribution of  
magnetic field strengths in MS stars respectively \citep[][and Shultz et al. in prep.]{Shultz2016}.
The predicted $\left(\frac{R_{\rm ZAMS}}{R_{\rm current}}\right)^2$ for our sample is between 10 and 
100 which means we expect to detect dipole fields of between 7\,G and 140\,G. 

We do not, however, measure $B_{\rm d}$. Instead,  we measure the mean longitudinal 
field $B_{\rm l}$ which is a function of the inclination angle ($i$) of the star and the obliquity angle ($\beta$) between the rotation axis and the magnetic axis \citep{Preston1967}. For our stars,  $i$ and $\beta$ were not available prior to the observations and so we use the equations from \citet{Preston1967} to estimate that conservatively $B_{\rm l}$ is $\sim$3 times weaker than $B_{\rm d}$. This lower value is factored 
into our calculations of the total required exposure time for each star.

\subsection{Observations}
\label{sec:obs}

The LIFE observations we present in this paper were taken using the 
ESPaDOnS (an Echelle SpectroPolarimetric Device 
for the Observation of Stars) instrument at the Canada France Hawaii Telescope (CFHT) 
and are summarised in Table~\ref{Table:LIFE_Targets}. We observe  over 
a wavelength range from about 3700\,\AA\ to 10500\,\AA\ with a resolving power of $\sim$\,68000. 
Each spectrum is spread over 40 echelle orders. The observations are corrected for the bias and flat-field, and a ThAr spectrum is used to 
calibrate the wavelengths to the pixel values. 
The data were reduced using {\sc Libre-Esprit} \citep{Donati1999} and {\sc Upena}, 
a software pipeline available at the CFHT. 

 The data were taken in circular
 spectropolarimetric mode, measuring Stokes $I$ and $V$. Each observation sequence 
consists of four sub-exposures, taken with different orientations of the Fresnel rhombs. The observations were constructively combined to form 
Stokes $V$ and destructively combined to form the null spectra, $N$ \citep{Donati1997}, using the ratio method \citep{Bagnulo2009}. 
Adding all four observations together provides Stokes $I$. If the total exposure time calculated 
in Section~\ref{sec:LIFE} was expected to saturate the detector, we took a number of 
consecutive sequences of the four sub-exposures which add up to the total required 
exposure time. These observations are then co-added after we have produced each least-squared deconvolution (LSD) profile (see Section~\ref{sec:LSD}).

\subsection{Normalisation of stellar spectra}

We used 
 a semi-automatic {\sc Python} program \citep[][and Appendix \ref{appendix:normal}]{Martin2017} to determine 
the continuum shape of the Stokes $I$ spectrum, fitting each ESPaDOnS order individually. 
The {\sc Python} program allows the user to fit a third-order spline to the continuum. 
This is achieved by initially fitting a third-order spline to the reduced unnormalised 
spectrum. Points from the spectrum are then iteratively $\upsigma$-clipped  
about the spline, until only continuum points remain. For the majority of spectra 
the $\upsigma$-clipping is asymmetric with more points clipped below than above the fit (e.g., $\sigma_{\rm upper}$ = 3 
and $\sigma_{\rm lower}$ = 1 ). 
This is because, to obtain a good continuum fit, all spectral lines must be removed and 
for the stars in this sample we mainly observe absorption lines. 
However, each parameter which determines the final continuum fit can be changed interactively 
including the number of iterations, the number of knots defining the cubic spline, and the 
$\upsigma$-clipping bounds. 
Finally, we take the best continuum model calculated for each star and use it 
to normalise the Stokes $I$ and $V$ and $N$ spectra.

\subsection{Least Squares Deconvolution}
\label{sec:LSD}
During this project we are searching for very weak magnetic fields, between $\sim$1\, and 100\,G. 
This leads to Stokes $V/I_c$ signatures with very low amplitudes. To detect such tiny signatures, 
we use the LSD technique \citep{Donati1997} and combine multiple consecutive observations
to increase the S/N of our data. We calculate 
mean LSD Stokes $I$, Stokes $V$ and $N$ profiles and co-add consecutive observations to 
produce the LSD line profiles which we use to measure the magnetic field of our observed stars. 
Since we aim to reduce all sources of noise  we must 
take care when producing the line mask. The line mask is used to determine which 
lines to include in the LSD profile and to provide the LSD routine with the 
parameters of each line including their wavelength, relative depth, and Land\'{e} factor. 
To start, we extract a stellar line list from the VALD3 database \citep{Piskunov1995,Kupka1999}. 
This line list is calculated with the $T_{\rm eff}$ and $\log g$ which we determine for each of 
our stars (see Section~\ref{sec:fund}). However, it will contain lines which are not present in the stellar spectrum 
and other lines which will add noise to the LSD profile. Therefore, we comb through each list, 
first removing lines with a depth smaller than 0.01 (relative to a continuum level of 1), and those which we do not 
see in our observed spectra. We remove hydrogen lines, because their shape is 
different from metal lines, and we remove any lines which blend with 
 either H, interstellar, or strong telluric lines. 
Finally, we adjust the  depth of each remaining line in the line mask so that it is 
consistent with the observed spectrum, in the manner described by \citet{Grunhut2017}.

\section{Stellar properties}
\label{sec:stelparams}

\begin{table*}
	\caption{Fundamental parameters of the observed stars. The name and spectral type of each star is given in the first two columns with references at the bottom of the table. The remaining columns show $T_{\rm eff}$, $\log g$, $v \sin i$ and $v_{\rm rad}$, which were calculated according to the methods described in Section~\ref{sec:fund}.}        
	\label{Table:Fundamental} 
	\centering                               
\begin{tabular}{l c r@{ $\pm$ }l  r@{ $\pm$ }l  r@{ $\pm$ }l r@{ $\pm$ }l r}       
		\hline\hline
		Star	& Spectral & \multicolumn{2}{c}{$T_{\rm eff}$} & \multicolumn{2}{c}{$\log g$}	
                           & \multicolumn{2}{c}{$v \sin i$}    & \multicolumn{2}{c}{$v_{\rm rad}$}\\
        		& Type     & \multicolumn{2}{c}{(K)}           & \multicolumn{2}{c}{(cgs)}
                           & \multicolumn{2}{c}{(km\,s$^{-1}$)}& \multicolumn{2}{c}{(km\,s$^{-1}$)}\\
        \hline 
             13\,Mon	&  A1Ib$^1$ 		&   10250 	&  300 	&      2.2 	& 0.2 	&13.4	& 0.1	&13.21	& 0.02\\	
             15\,Sgr	&O9.7Iab$^2$		&   30000 	&  1000 &      3.5 	& 0.2 	&58		& 1		&19.8	& 0.1\\		
             19\,Aur	&A5Ib-II$^3$		&    8500 	&  200 	&      2.0 	& 0.2 	&8.8	& 0.1	&-3.15	& 0.01\\	
             24\,CMa	&  B4Ia$^1$ 		&   17000 	&  400 	&      2.1 	& 0.2 	&39.2	& 0.4	&48.08	& 0.07\\	
       $\upeta$\,Leo	&  A0Ib$^1$ 		&    9750 	&  300 	&      2.0 	& 0.2 	&13.9	& 0.1	&2.82	& 0.02\\	
     $\upgamma$\,CMa	& B6III$^1$ 		&   13600 	&  300 	&      3.4 	& 0.2 	&37		& 1		&31.1	& 0.2\\		
           HD\,10362	& B7III$^4$ 		&   14300 	&  300 	&      3.3 	& 0.2 	&30.1	& 0.7	&-6.0	& 0.1\\		
           HD\,42035	&  B9V$^5$  		&   10500 	&  200 	&      3.5 	& 0.2 	&4		& 1		&2.3	& 0.02\\	
          HD\,186660	&  B2III/IV$^6$  		&   16900 	&  300 	&      3.6 	& 0.2 	&11.4	& 0.1	&-16.65	& 0.02\\	
          HD\,188209	&O9.5Iab$^7$		&   30000 	&  1000 &      3.1 	& 0.2 	&84		& 1		&-18.78	& 0.08\\	
          HD\,209419	&  B5V$^8$  		&   14100 	&  300 	&      3.5 	& 0.2 	&14.5	& 0.3	&-16.23	& 0.05\\	
          HIP\,38584	&B8II$^9$			&   12600 	&  300 	&      3.0 	& 0.2 	&28.4	& 0.6	&30.0		& 0.1\\
            HR\,3042	& B8/9II$^{10}$ 	&   14150 	&  300 	&      3.5 	& 0.2 	&60		& 1		&-4.5		& 0.2\\
             PT\,Pup	& B1V$^{11}$		&   26300 	&  500 	&      4.1 	& 0.2 	&16.9	& 0.1	&37.83	& 0.02\\
           V399\,Lac	&  A3Ib$^3$ 		&    8500 	&  200 	&      1.5 	& 0.2 	&11.8	& 0.1	&-25.30	& 0.01\\
         \hline
		\multicolumn{11}{l}{Note: taken from $^1$\citet{Zorec2009}, $^2$\citet{Sota2014}, $^3$\citet{Abt1995}, }\\
		\multicolumn{11}{l}{$^4$\citet{Jensen1981}, $^5$\citet{Molnar1972}, $^6$\citet{Houk1999}, $^7$\citet{Sota2011}, }\\
		\multicolumn{11}{l}{$^8$\citet{Gkouvelis2016}, $^{9}$\citet{Houk1978}, $^{10}$\citet{Houk1988}, $^{11}$\citet{Nieva2013}.}\\
		          	                                  
	\end{tabular}
\end{table*}
\begin{figure}
  \includegraphics[width=\columnwidth]{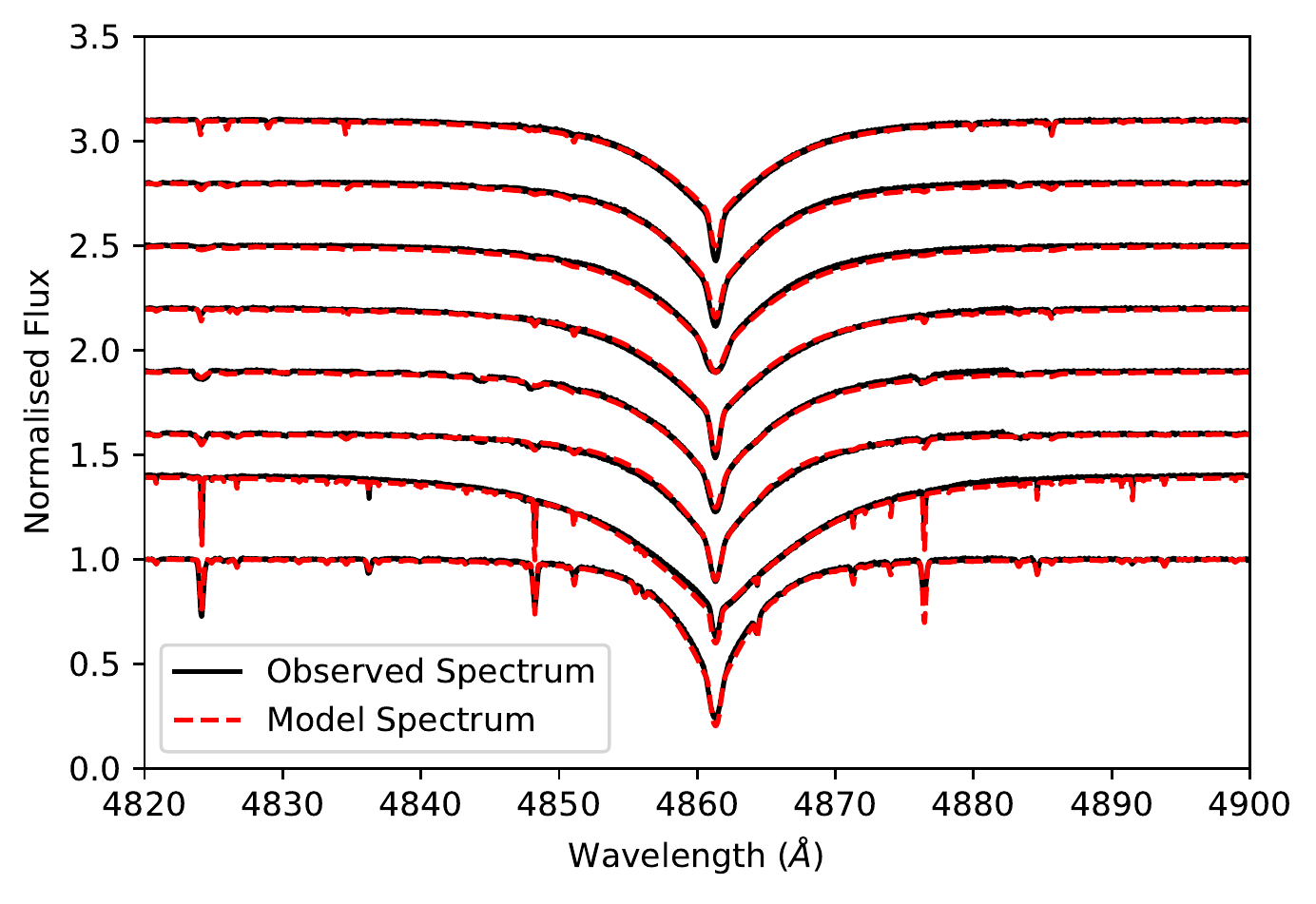}
  \caption{The observed H$\upbeta$ lines (black solid lines) and corresponding
 model spectra (red dashed line). The stars shown are those without 
significant wind contributions. From top to bottom
     the stars are: HD\,186660, HD\,10362, HR\,3042, HD\,209419, $\upgamma$CMa, HIP\,38584, HD\,42035 and 13\,Mon. Each profile is calculated 
     with fundamental parameters as shown in Table~\ref{Table:Fundamental}.}
  \label{fig:Balmer}
\end{figure}
\subsection{Fundamental parameters}
\label{sec:fund}
For each star we calculated the effective temperature ($T_{\rm eff}$) and 
surface gravity ($\log g$) using the 
{\sc uvbybeta} code \citep{Napiwotzki1993}. We also used grids
of synthetic spectra to visually check that the values accurately reproduce primarily the H$\upalpha$, H$\upbeta$, 
H$\upgamma$ and H$\updelta$ lines, but also regions of metal lines. Examples of the fit between the synthetic and observed spectrum  for H$\upbeta$ are shown in Fig.~\ref{fig:Balmer}. 
 The 
grids of synthetic spectra we used were 
calculated by \citet[][using {\sc atlas9} model atmospheres]{Bohlin2017}, 
\citet[][using {\sc atlas9} model atmospheres]{Martin2017} and 
\citet[][using {\sc tlusty} model atmospheres]{Lanz2007}. In the case of {\sc atlas9} 
model atmospheres \citep{Kurucz1993}, the synthetic 
spectra are calculated assuming plane parallel geometry, local thermodynamic equilibrium (LTE) 
and opacity distribution functions (ODFs) for solar abundances \citep{Kurucz1993b}. 
The synthetic spectra were computed with {\sc cossam\_simple} \citep{Martin2017} and 
 {\sc synthe} \citep{Kurucz1981,Bohlin2017}. For the {\sc tlusty} 
model atmospheres the synthetic spectra are calculated assuming 
non-local thermodynamic equilibrium, plane-parallel geometry and hydrostatic
equilibrium using {\sc synspec} \citep{Hubeny2011}. Where necessary we calculated additional synthetic spectra using {\sc atlas9} 
model atmospheres and  {\sc synspec}.

To calculate the projected rotational velocity ($v \sin i$) and the radial velocity ($v_{\rm rad}$) for each star, we first fit a 
Gaussian to the LSD profile. The mean value of this Gaussian is adopted as the 
 $v_{\rm rad}$ of the star. We then subtract $v_{\rm rad}$ from the velocity of each pixel such that
 the line center of the LSD profile is shifted to 0\,kms$^{-1}$.
We apply a Fast Fourier transform (FFT) to the profile. Following 
 \citet{Gray2005} and \citet{Glazunova2008}, the first minimum of the 
transform is related to the stellar $v \sin i$ value by the limb darkening 
coefficient. The limb darkening value varies as a function of stellar $T_{\rm eff}$ and 
$\log g$;  for each of our stars we use limb darkening values from 
\citet{Claret2011} which are consistent with our calculated $T_{\rm eff}$ and $\log g$ values.

We calculate the minimum error on $v_{\rm rad}$ ($\sigma_{v_{\rm rad}}$) following \citet{Seager2010} as
\begin{equation}
\sigma_{v_{\rm rad}} = \frac{\sqrt{{\rm FWHM_{lsd}}}}{I\cdot {\rm SNR}},
\end{equation}
where ${\rm FWHM_{lsd}}$ is the full width half maximum of the 
LSD profile and $I$ is the maximum intensity value of the inverse of the LSD profile.
 We calculate 
the error on $v\sin i$ ($\sigma_{v\sin i}$)  following \citet{Diaz2011} 
as
\begin{equation}
\sigma_{v\sin i} = 4.42 \cdot {\rm FWHM_{lsd}}^{0.520} \cdot {\rm rms} \cdot I^{-1.08},
\end{equation}
where  rms is the  root mean square of the LSD profile.

To check the consistency of the calculated  
$v \sin i$  and $v_{\rm rad}$ values, we compare the observed spectrum with a 
synthetic spectrum calculated using our derived fundamental parameters and we find that they agree well.\\
\begin{table*}
	\caption{Stellar parameters of the observed stars. The parallax is taken from 
    the Hipparcos archive \citep{vanLeeuwen2007} or where available the GAIA data release 1  \citep[denoted by *;][]{Gaia2016a,Gaia2016b}.
    The visual magnitude is taken from the Hipparcos archive \citep{Perryman1997} and the Bolometric correction is calculated 
following  \citet{Flower1996} and \citet{Torres2010} using the $T_{\rm eff}$  shown in Table~\ref{Table:Fundamental}. The columns $L$, $\mathcal{L}$, M$_{\rm ZAMS}$, $\upOmega_{\rm ZAMS}$, R$_{\rm ZAMS}$, M, R, Age and Turn-off Age are determined as described in Section~\ref{sec:luminosity}. Bold values of Age show those less than the turn-off age, therefore suggesting the star may still be on the MS. The final column shows the expected dipolar magnetic field strength for each star if the star hosted a 0.63\,kG dipolar field at the ZAMS and flux conservation is the main contributing factor to the reduction of  surface field strength. }
	\label{Table:Lumino_Params} 
	\centering                               
	\resizebox{\textwidth}{!}{
    \begin{tabular}{l r@{\,$\pm$\,}lcccr@{\,$\pm$\,}lcccccccc}   
		\hline \hline
		Star &	\multicolumn{2}{c}{Parallax} & $M_V$	& Bolometric & $\log L$ & \multicolumn{2}{c}{$\log \mathcal{L}$} 
        & M$_{\rm ZAMS}$			         & $\upOmega_{\rm ZAMS}$			
        & R$_{\rm ZAMS}$			         & M
		& R 								 & Age 
        & Turn-off & Exp. Field\\   
        & \multicolumn{2}{c}{(mas)}      &          & correction
                         & ($\log  L_{\odot}$)	& \multicolumn{2}{c}{($\log \mathcal{L}_{\odot}$)}
						 & (M$_{\odot}$)				         & ($\upOmega_{\rm crit}$)			
                         & (R$_{\odot}$)                       & (M$_{\odot}$)         
                         & (R$_{\odot}$)		                 & (Myr)                             
                         &  Age (Myr) & (G)\\   
		\hline   
             13\,Mon	&0.8	&0.5	&-6.1$^{+1.0}_{-1.8}$	& -0.30	&4.4$^{+0.4}_{-0.7}$	&3.2 & 0.1	&15.0	& 0.60	& 4.7	& 14.7	& 91.9	& 13.8	& 13.6	&    1\\
	&\multicolumn{2}{c}{}	&	&	&	&\multicolumn{2}{c}{}	& 8.0	& 0.00	& 3.2	& 8.0	& 23.8	& 33.3	& 32.9	&   11\\
             15\,Sgr	&0.1	&0.5	&-10.0\,$\pm$\,10.0	& -2.89	&7\,$\pm$\,4	&3.79 & 0.09	&32.0	& 0.00	& 6.9	& 30.3	& 18.1	& {\bf 4.8}	& 5.2	&   92\\
	&\multicolumn{2}{c}{}	&	&	&	&\multicolumn{2}{c}{}	& 23.0	& 0.00	& 5.7	& 22.5	& 12.3	& {\bf 5.9}	& 6.8	&  140\\
             19\,Aur	&1.6	&0.3	&-4.4$^{+0.4}_{-0.5}$	& -0.00	&3.7\,$\pm$\,0.2	&3.1 & 0.1	&10.0	& 0.00	& 3.6	& 9.7	& 59.3	& 23.8	& 20.9	&    2\\
	&\multicolumn{2}{c}{}	&	&	&	&\multicolumn{2}{c}{}	& 7.0	& 0.60	& 3.1	& 6.9	& 36.8	& 56.7	& 51.0	&    4\\
             24\,CMa	&1.2	&0.4	&-6.7$^{+0.6}_{-0.9}$	& -1.53	&5.2$^{+0.3}_{-0.4}$	&4.2 & 0.1	&40.0	& 0.00	& 7.8	& 16.1	& 71.6	& 4.6	& 4.5	&    7\\
	&\multicolumn{2}{c}{}	&	&	&	&\multicolumn{2}{c}{}	& 15.0	& 0.59	& 4.7	& 14.7	& 29.4	& 13.6	& 13.5	&   16\\
	  $\upeta$\,Leo	&2.6	&0.2	&-4.5\,$\pm$\,0.1	& -0.20	&3.77$^{+0.05}_{-0.06}$	&3.3 & 0.2	&12.0	& 0.98	& 4.8	& 11.9	& 66.6	& 20.3	& 19.2	&    3\\
	&\multicolumn{2}{c}{}	&	&	&	&\multicolumn{2}{c}{}	& 8.0	& 0.00	& 3.2	& 8.0	& 25.9	& 33.3	& 32.9	&    9\\
      $\upgamma$\,CMa	&7.4	&0.2	&-1.49\,$\pm$\,0.06	& -1.00	&2.88$^{+0.02}_{-0.03}$	&2.5 & 0.1	&7.0	& 0.00	& 2.9	& 7.0	& 11.5	& 42.3	& 41.8	&   41\\
	&\multicolumn{2}{c}{}	&	&	&	&\multicolumn{2}{c}{}	& 5.0	& 0.98	& 2.9	& 5.0	& 5.5	& {\bf 98.9}	& 113.6	&  170\\
           HD\,10362	&2.4	&0.5*	&-2.2$^{+0.4}_{-0.5}$	& -1.12	&3.2\,$\pm$\,0.2	&2.7 & 0.1	&7.0	& 0.11	& 3.0	& 7.0	& 10.7	& 45.2	& 44.7	&   48\\
	&\multicolumn{2}{c}{}	&	&	&	&\multicolumn{2}{c}{}	& 5.0	& 0.33	& 2.5	& 5.0	& 5.3	& {\bf 97.8}	& 104.5	&  140\\
          HD\,186660	&2.1	&0.5*	&-2.7$^{+0.5}_{-0.6}$	& -1.52	&3.6\,$\pm$\,0.2	&2.70 & 0.09	&8.0	& 0.00	& 3.2	& 8.0	& 9.4	& 33.2	& 32.9	&   72\\
	&\multicolumn{2}{c}{}	&	&	&	&\multicolumn{2}{c}{}	& 7.0	& 0.97	& 3.5	& 7.0	& 6.5	& {\bf 46.5}	& 52.8	&  180\\
          HD\,188209	&0.9	&0.2	&-5.4$^{+0.5}_{-0.6}$	& -2.89	&5.2\,$\pm$\,0.2	&4.19 & 0.09	&84.9	& 0.00	& 12.2	& 70.0	& 40.9	& {\bf 2.5}	& 3.0	&   56\\
	&\multicolumn{2}{c}{}	&	&	&	&\multicolumn{2}{c}{}	& 20.0	& 0.57	& 5.5	& 19.7	& 10.7	& {\bf 7.6}	& 9.5	&  170\\
          HD\,209419	&2.9	&0.5*	&-2.1\,$\pm$\,0.4	& -1.09	&3.2$^{+0.1}_{-0.2}$	&2.5 & 0.1	&5.0	& 0.54	& 2.5	& 5.0	& 6.4	& 110.1	& 109.2	&   98\\
	&\multicolumn{2}{c}{}	&	&	&	&\multicolumn{2}{c}{}	& 5.0	& 0.00	& 2.4	& 5.0	& 5.2	& {\bf 88.1}	& 88.2	&  130\\
          HIP\,38584	&1.5	&0.4	&-2.2$^{+0.5}_{-0.7}$	& -0.81	&3.1$^{+0.2}_{-0.3}$	&2.8 & 0.1	&8.0	& 0.00	& 3.2	& 8.0	& 16.6	& 33.3	& 32.9	&   23\\
	&\multicolumn{2}{c}{}	&	&	&	&\multicolumn{2}{c}{}	& 4.0	& 0.58	& 2.2	& 4.0	& 5.1	& 191.0	& 189.4&  120\\
            HR\,3042	&2.8	&0.6*	&-1.5$^{+0.4}_{-0.5}$	& -1.10	&2.9\,$\pm$\,0.2	&2.5 & 0.1	&5.0	& 0.54	& 2.5	& 5.0	& 6.3	& 110.1	& 109.2	&   99\\
	&\multicolumn{2}{c}{}	&	&	&	&\multicolumn{2}{c}{}	& 5.0	& 0.98	& 2.9	& 5.0	& 4.9	& {\bf 91.7}	& 113.6	&  220\\
             PT\,Pup	&1.9	&0.3	&-3.1\,$\pm$\,0.3	& -2.53	&4.1\,$\pm$\,0.1	&2.96 & 0.07	&12.0	& 0.73	& 4.2	& 12.0	& 5.4	& {\bf 9.1}	& 18.9	&  380\\
	&\multicolumn{2}{c}{}	&	&	&	&\multicolumn{2}{c}{}	& 11.0	& 0.00	& 3.8	& 11.0	& 4.5	& {\bf 6.3}	& 17.7	&  450\\
           V399\,Lac	&0.5	&0.3	&-6.7$^{+1.0}_{-1.8}$	& -0.00	&4.6$^{+0.4}_{-0.7}$	&3.6 & 0.2	&15.0	& 0.95	& 5.2	& 14.4	& 129.7	& 15.1	& 14.2	&    1\\
	&\multicolumn{2}{c}{}	&	&	&	&\multicolumn{2}{c}{}	& 7.0	& 0.64	& 3.1	& 6.9	& 38.7	& 58.9	& 51.7	&    4\\

		\hline
	\end{tabular}}
\end{table*}

\subsection{Luminosity and evolutionary status}
\label{sec:luminosity}

To determine the evolutionary status of each star, along with its radius ($R$) and mass ($M$), we 
use the luminosity, $L$, and the quantity $\mathcal{L}= T^4_{\rm eff}/g$. We calculate $\log L/L_{\rm \odot}$
using parallaxes from the Hipparcos database \citep{vanLeeuwen2007} or where available the GAIA data release 1  \citep{Gaia2016a,Gaia2016b} and the Johnson visual magnitudes given by  \citet{Perryman1997}. We calculate the 
bolometric correction for each star following \citet{Flower1996} and \citet{Torres2010} using the temperatures given in Table~\ref{Table:Fundamental}. 

\begin{figure*}
 \includegraphics[width=0.5\textwidth,height=.89\textheight]{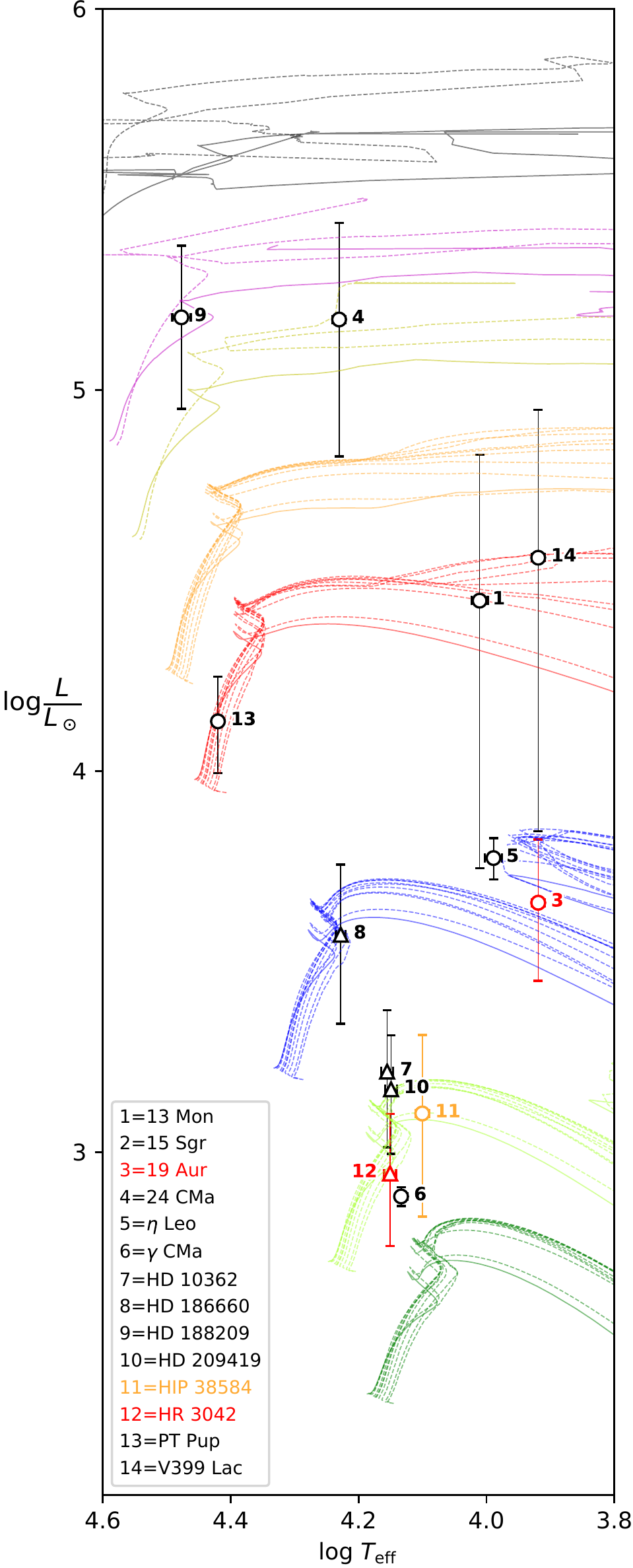}\includegraphics[width=0.5\textwidth,height=.89\textheight]{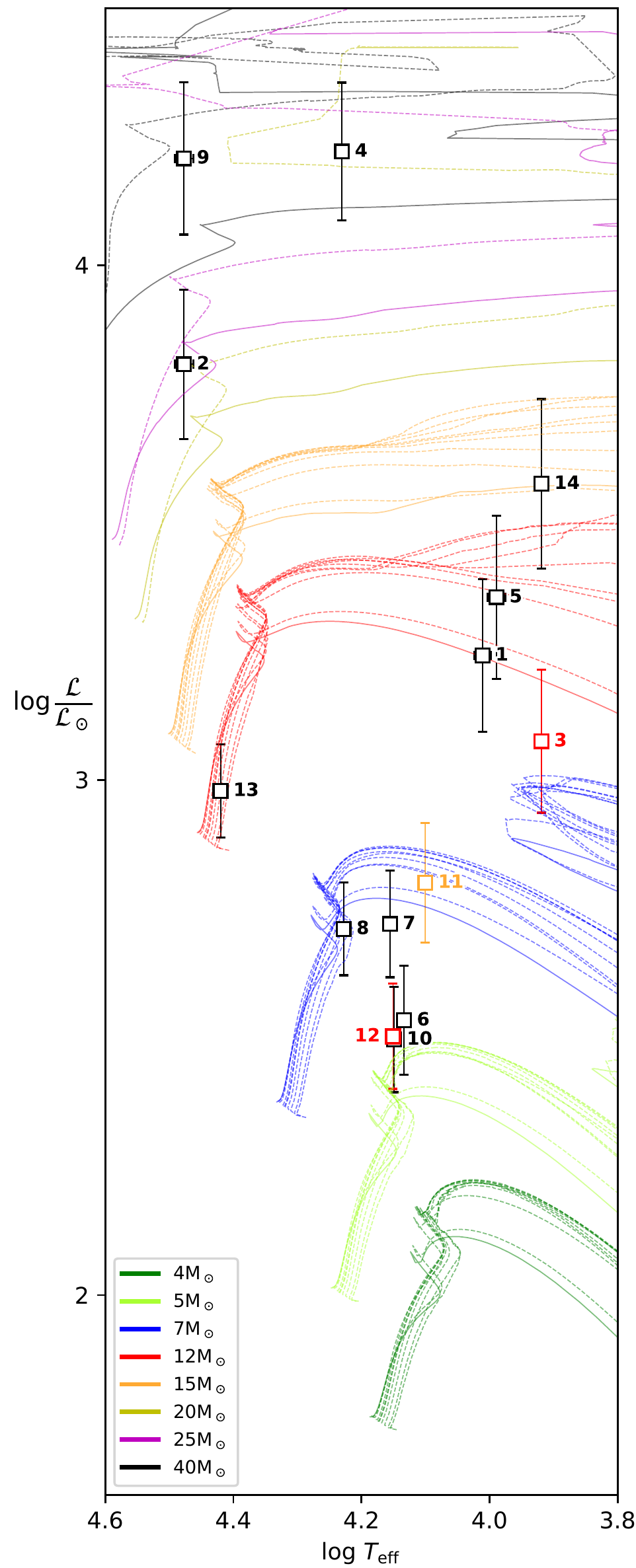}
  \caption{Left panel: Hertzsprung-Russell (HR) diagram of the LIFE targets, the circles are luminosities calculated with Hipparcos parallaxes \citep{vanLeeuwen2007} and the triangles are with GAIA parallaxes \citep{Gaia2016a,Gaia2016b}.  Right panel: Spectroscopic HR diagram of the LIFE targets, where we calculate $\log (\mathcal{L}/\mathcal{L}_{\odot})$ following \citet{Langer2014} shown with squares. In both panels, the  evolutionary tracks  from \citet{Ekstrom2012} and \citet{Georgy2013} are plotted (solid line, no rotation and dashed lines with rotation from $\upOmega_{\rm ZAMS}/\upOmega_{\rm crit}$=0.1-1.0; the colour of the isochrones denotes its mass). Red points are those stars we find to be magnetic, the orange point in each panel is the magnetic candidate and black points are non-magnetic stars. The number closest to each point identifies the corresponding star. The errorbars for $\log T_{\rm eff}$ are plotted but, in general, they are smaller than the  symbol used to indicate each star.}
  \label{fig:Hr}
\end{figure*}

We calculate $\mathcal{L}/\mathcal{L}_{\rm \odot}$ 
following \citet{Langer2014}, 
\begin{equation}
\log \frac{\mathcal{L}}{\mathcal{L}_{\rm \odot}} = 	\log\left(\frac{T_{\rm eff}^4}{g}\right)- 
								\log\left(\frac{T_{\rm eff\odot}^4}{g_\odot}\right),
\end{equation}
where $T_{\rm eff}$ and $\log g$ are given in Table~\ref{Table:Fundamental}, 
$T_{\rm eff\odot}$ is the solar 
effective temperature (5756K) and $g_\odot$ is the solar surface gravity (27542 cgs) used as calibration values for the \citet{Ekstrom2012} and \citet{Georgy2013} evolutionary tracks. 
These values are given in Table~\ref{Table:Lumino_Params}. For each star, we plot   
$\log L/L{\rm \odot}$ on the theoretical Hertzsprung-Russell (HR) diagram and 
$\log \mathcal{L}/\mathcal{L}{\rm \odot}$ on the spectroscopic HR diagram shown in the 
left and right panels of
Fig. \ref{fig:Hr} respectively. 
In addition, we plot theoretical evolutionary tracks 
which take into account the effects of rotation, but not the effects of  magnetic fields
\citep{Ekstrom2012,Georgy2013}. \\

The evolutionary tracks  from \citet{Ekstrom2012} 
 cover a range of ZAMS masses ($M_{\textsc{ZAMS}}$) from 
0.8 to 120 M$_\odot$ and fractional critical velocities\footnote{The critical velocity of a star describes the stellar rotational velocity at which the centrifugal force at the equator balances with the gravitational force.}  at the ZAMS ($\upOmega_{\rm ZAMS}/\upOmega_{\rm crit}$) of  0.0 and 0.4. The models from \citet{Georgy2013}  cover a range of ZAMS masses ($M_{\textsc{ZAMS}}$) from 
1.7 to 15 M$_\odot$ and fractional critical velocities at the ZAMS ($\upOmega_{\rm ZAMS}/\upOmega_{\rm crit}$) from 0.0 to 1.0. All models used are computed with solar metallicity (Z = 0.014).

For each evolutionary track a variety of stellar parameters are provided as a function of stellar age. For this study we are particularly interested in the ratio $R/R_{\rm ZAMS}$ and the comparison between the age of the star, $t$,  and the predicted age at which the star turns off the main sequence. To infer these values for the stars in this study we identify the models which minimise the following expressions:

\begin{equation*}
\begin{aligned}
1.) \underset{M_\textsc{zams}, \upOmega_\textsc{zams}, t}{\text{minimize}} 
&\left\{\left[\frac{L_{\rm evo}}{L_{\rm \odot}}\left(M_\textsc{zams},\upOmega_\textsc{zams},t\right) -  \left( \frac{L_*}{L_{\rm \odot}} + \sigma_{\frac{L_*}{L_{\rm \odot}}}\right)\right]^2 + \right.\\  
&\left.\left[T_{\rm eff, evo}\left(M_\textsc{zams},\upOmega_\textsc{zams},t\right) - \left(T_{\rm eff,*} + \sigma_{T_{\rm eff,*}}\right)\right]^2\right\}^{0.5} 
\end{aligned}
\end{equation*}
\vspace{-11pt}
\begin{equation*}
\begin{aligned}
2.) \textrm{ Same as 1 but for} \log \mathcal{L}/\mathcal{L}{\rm \odot}\\
\end{aligned}
\end{equation*}
\vspace{-11pt}
\begin{equation*}
\begin{aligned}
3.) \underset{M_\textsc{zams}, \upOmega_\textsc{zams}, t}{\text{minimize}} 
&\left\{\left[\frac{L_{\rm evo}}{L_{\rm \odot}}\left(M_\textsc{zams},\upOmega_\textsc{zams},t\right) -  \left( \frac{L_*}{L_{\rm \odot}} - \sigma_{\frac{L_*}{L_{\rm \odot}}}\right)\right]^2 + \right.\\  
&\left.\left[T_{\rm eff, evo}\left(M_\textsc{zams},\upOmega_\textsc{zams},t\right) - \left(T_{\rm eff,*} - \sigma_{T_{\rm eff,*}}\right)\right]^2\right\}^{0.5} 
\end{aligned}
\end{equation*}
\vspace{-11pt}
\begin{equation*}
\begin{aligned}
4.) \textrm{ Same as 3 but for} \log \mathcal{L}/\mathcal{L}{\rm \odot}.\\
\end{aligned}
\end{equation*}
The subscripts evo and * represent values taken from the evolutionary model and the stellar data respectively. 

In Table~\ref{Table:Lumino_Params}, we give the $M_{\rm ZAMS}$, $\upOmega_{\rm ZAMS}$, $R_{\rm ZAMS}$, $M$, $R$, age and turn-off age for each star, taken from two of the above models. Specifically, the first row for each star reports the parameter set described by the model 1 or 2 which maximises $M_\textsc{zams}\textrm{, } \upOmega_\textsc{zams}, \textrm{ and } t$. The second row reports the parameter set described by the model 3 or 4 which minimises $M_\textsc{zams}\textrm{, } \upOmega_\textsc{zams}, \textrm{ and } t$. The consequence of using this method is in some cases the two reported sets of values for a star may be considerably different. However, this is done in order to be conservative.

In the following subsections we comment on our findings for each of the observed stars.

\subsection{13\,Mon}
For 13\,Mon we determine the $T_{\rm eff}$  to be $10250\pm300\,$K and the 
 $\log g$ to be $2.2\pm0.2$. These values agree with those calculated 
by \citet[][$T_{\rm eff}  = 10000\pm200\,$K and $\log g = 2.15\pm0.1$]{Firnstein2012} but these values are slightly higher than what would be expected from a star with the  
spectral type of A1Ib suggested by \citet{Zorec2009}. However, the plots in Fig.~\ref{fig:Hr} show  
that 13\,Mon is very clearly post-MS. We calculate an age between 13.8 and 33.3\,Myr, which can be  compared to the age at the MS turn-off of between 13.6 and 32.9\,Myr, respectively. We conclude that it has a mass of 8--15\,M$_{\odot}$ and a radius of 23.8--91.9\,R$_{\odot}$. This suggests a radius increase of between a factor of 7.5 and 19.5 since the ZAMS.
\begin{figure}
  \includegraphics[width=\columnwidth]{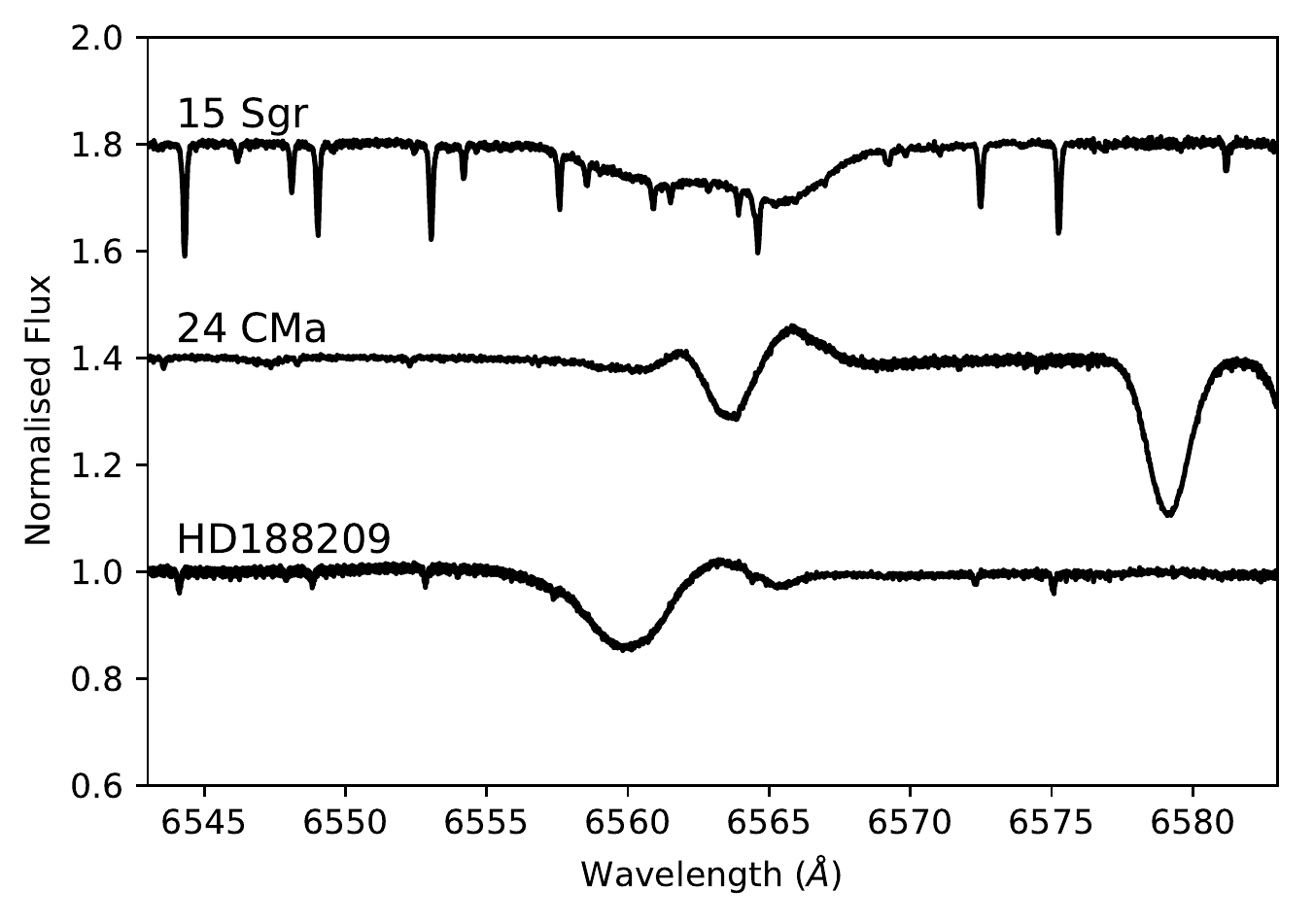}
  \caption{The observed H$\upalpha$ lines of (from top to bottom): 15\,Sgr and 24\,CMa and HD\,188209.}
  \label{fig:alpha}
\end{figure}
\subsection{15\,Sgr}
\label{sec:15sgr}
Analyses of 15\,Sgr by \citet{Sana2014}, \citet{Sota2014} and \citet{Tokovinin2010} 
have shown that it is a binary system. However, we do not see the presence of a companion in our spectra, likely because 
the magnitude of the secondary is much lower than the primary. We do see the presence of emission in H$\upalpha$ (see Fig.~\ref{fig:alpha}) 
suggesting a stellar wind, consistent with this star being an O supergiant.

Since 15\,Sgr is an O supergiant, our determination of $T_{\rm eff}$ and $\log g$ is focussed mainly on the analysis with {\sc uvbybeta}, 
in addition to the results of previous studies. This is because our synthetic spectra assume a plane parallel atmosphere. Even so, we obtain good agreement between our synthetic spectrum and the observed spectrum. The spectral type of 09.7Iab \citep{Sota2014} and the fundamental parameters determined 
by \citet[][$T_{\rm eff}  =  31500\,{\rm K}$ and $\log g = 3.5$]{Trundle2002} and 
by \citet[][$T_{\rm eff}  = 30000\,$K and $\log g = 3.5$]{Grunhut2017} 
agree with our findings of $T_{\rm eff}$ ($30000\pm1000\,$K) and  
 $\log g$ ($3.5\pm0.2$). The uncertainity of the parallax measurement of 15\,Sgr is very large, which leads to an inconclusive luminosity value. As a result, for this star, we only plot $\mathcal{L}$ in Fig.~\ref{fig:Hr}. Figure~\ref{fig:Hr} shows 
that 15\,Sgr is most likely still on the MS and we calculate an age of between 4.8 and 5.9\,Myr compared to the age at the MS turn-off of between 5.2 and 6.8\,Myr, respectively. We conclude that it has a mass of 22.5--30.3\,M$_{\odot}$ and a radius of 12.3--18.1\,R$_{\odot}$. This suggests a radius increase of between a factor of 2.1 and 2.6 since the ZAMS.
\subsection{19\,Aur}
The star 19 Aur has been studied by \citet{Lyubimkov2010}
who found $T_{\rm eff} = 8300\pm100$\,{\rm K} and $\log g = 2.1 \pm 0.25$. 
This is consistent with our results of $T_{\rm eff}$=$8500\pm 200\,$K 
and $\log g$=$2.0\pm0.2$. These values also agree well with the spectral type of A5Ib-II \citep{Abt1995}.  The plots in Fig.~\ref{fig:Hr} show  
that 19\,Aur is a post-MS star and we calculate an age of between 23.8 and 56.7\,Myr compared to the age at the MS turn-off of between 20.9 and 51.0\,Myr, respectively. We conclude that it has a mass of 6.9--9.7\,M$_{\odot}$ and a radius of 36.8--59.3\,R$_{\odot}$. This suggests a radius increase of between a factor of 12.0 and 16.5 since the ZAMS.
\subsection{24\,CMa}
The fundamental parameters determined by 
\citet[][$T_{\rm eff}  = 16500\pm500\,$K and $\log g = 2.25$]{Searle2008} 
and
by \citet[][$T_{\rm eff}  = 17000\,$K and $\log g = 2.15$]{Lefever2007} 
agree with our findings of $T_{\rm eff}$ ($17000\pm400\,$K) and  
 $\log g$ ($2.1\pm0.2$).
The parameters calculated by \citet[][$T_{\rm eff}  =  15400\,{\rm K}$ and $\log g = 2.15$]{Fraser2010} and 
by \citet[][$T_{\rm eff}  = 15500\,$K and $\log g = 2.05$]{Crowther2006} are 
lower for both $T_{\rm eff}$ and $\log g$. 
However, all of the calculated parameters and the spectral type of B4Ia \citep{Zorec2009} lead to the conclusion that 24\,CMa is most likely a 
 post-MS star or at the very end of the MS. 
The H$\upalpha$ line of
 24\,CMa has been shown to vary in strength by 66.4\% \citep{Morel2004}, which could 
provide an explanation for the spread of results for the fundamental parameters. 
Our analysis is based on the inspection of a number 
of lines, and so should be less affected by the variation of H$\upalpha$. H$\upalpha$ is in emission (see Fig.~\ref{fig:alpha}) which is common in Ia supergiants, but H$\upalpha$ variation can also point to the presence of a magnetosphere and hence of a magnetic field. Therefore this target is particularly interesting for a spectropolarimetric study.

The study by \citet{Helden1972b} found 24\,CMa to be O deficient, however 
\citet{Walborn1976} found a normal CNO spectrum for a star with this spectral 
classification. Our findings also suggest this star has a normal CNO spectrum.

We calculate that 24\,CMa  has an age of between 4.6 and 13.6\,Myr compared to the age at the MS turn-off of between 4.5 and 13.5\,Myr, respectively. We conclude that it has a mass of 14.7--16.1\,M$_{\odot}$ and a radius of 29.4--71.6\,R$_{\odot}$. This suggests a radius increase of between a factor of 6.3 and 9.1 since the ZAMS.
\subsection{$\upeta$\,Leo}
For $\upeta$\,Leo we determine the $T_{\rm eff}$  to be $9750\pm300\,$K and the 
 $\log g$ to be $2.0\pm0.2$. This agrees well with the spectral type A0Ib \citep{Zorec2009} and with the 
values calculated 
by \citet[][$T_{\rm eff}  = 9600\pm 200\,$K and $\log g = 2.05\pm0.10$]{Firnstein2012}. 
The plots in Fig.~\ref{fig:Hr} show 
that $\upeta$\,Leo is well in to the post-MS  and we calculate an age of between 20.3 and 33.3\,Myr compared to the age at the MS turn-off of between 19.2 and 32.9\,Myr, respectively. We conclude that it has a mass of 8.0--11.9\,M$_{\odot}$ and a radius of 25.9--66.6\,R$_{\odot}$. This suggests a radius increase of between a factor of 8.1 and 13.9 since the ZAMS.
\subsection{$\upgamma$\,CMa}
$\upgamma$\,CMa has been shown to be a HgMn star \citep{Schneider1981} and a spectroscopic binary 
\citep{Schneider1981,Scholler2010}. It shows weak Y lines \citep{Hubrig2012} and has a 
rotation period of  6.16 days \citep{Briquet2010}. The spectral type B6III \citep{Zorec2009} and the fundamental parameters 
determined by \citet[][$T_{\rm eff}  =  13596\,{\rm K}$]{Makaganiuk2011} and 
by \citet[][$T_{\rm eff}  = 13600\,{\rm K}$ and $\log g = 3.40$]{Ghazaryan2016} 
agree perfectly with our findings of $T_{\rm eff}=13600\pm 300\,$K  and
 $\log g=3.4\pm 0.2$. 
The plots in Fig.~\ref{fig:Hr} show  
that $\upgamma$\,CMa is either a MS star or at the very start of the post-MS. We calculate an age of between 42.3 and 98.9\,Myr compared to the age at the MS turn-off of between 41.8 and 113.6\,Myr, respectively. We conclude that it has a mass of 5.0--7.0\,M$_{\odot}$ and a radius of 5.5--11.5\,R$_{\odot}$. This suggests a radius increase of between a factor of 1.9 and 3.9 since the ZAMS.
\subsection{HD\,10362}
For HD\,10362  we determine the $T_{\rm eff}$  to be $14300\pm300\,$K and the 
 $\log g$ to be $3.3\pm 0.2$, which is consistent with the spectral type of B7III \citep{Jensen1981}. The plots in Fig.~\ref{fig:Hr} show  
that HD\,10362 is post-MS or very close to the end of the MS. We calculate an age of between 45.2 and 97.8\,Myr compared to the age at the MS turn-off of between 44.7 and 104.5\,Myr, respectively. We conclude that it has a mass of 5.0--7.0\,M$_{\odot}$ and a radius of 5.3--10.7\,R$_{\odot}$. This suggests a radius increase of between a factor of 2.1 and 3.6 since the ZAMS.
\subsection{HD\,42035}
\begin{figure}
  \includegraphics[width=\columnwidth]{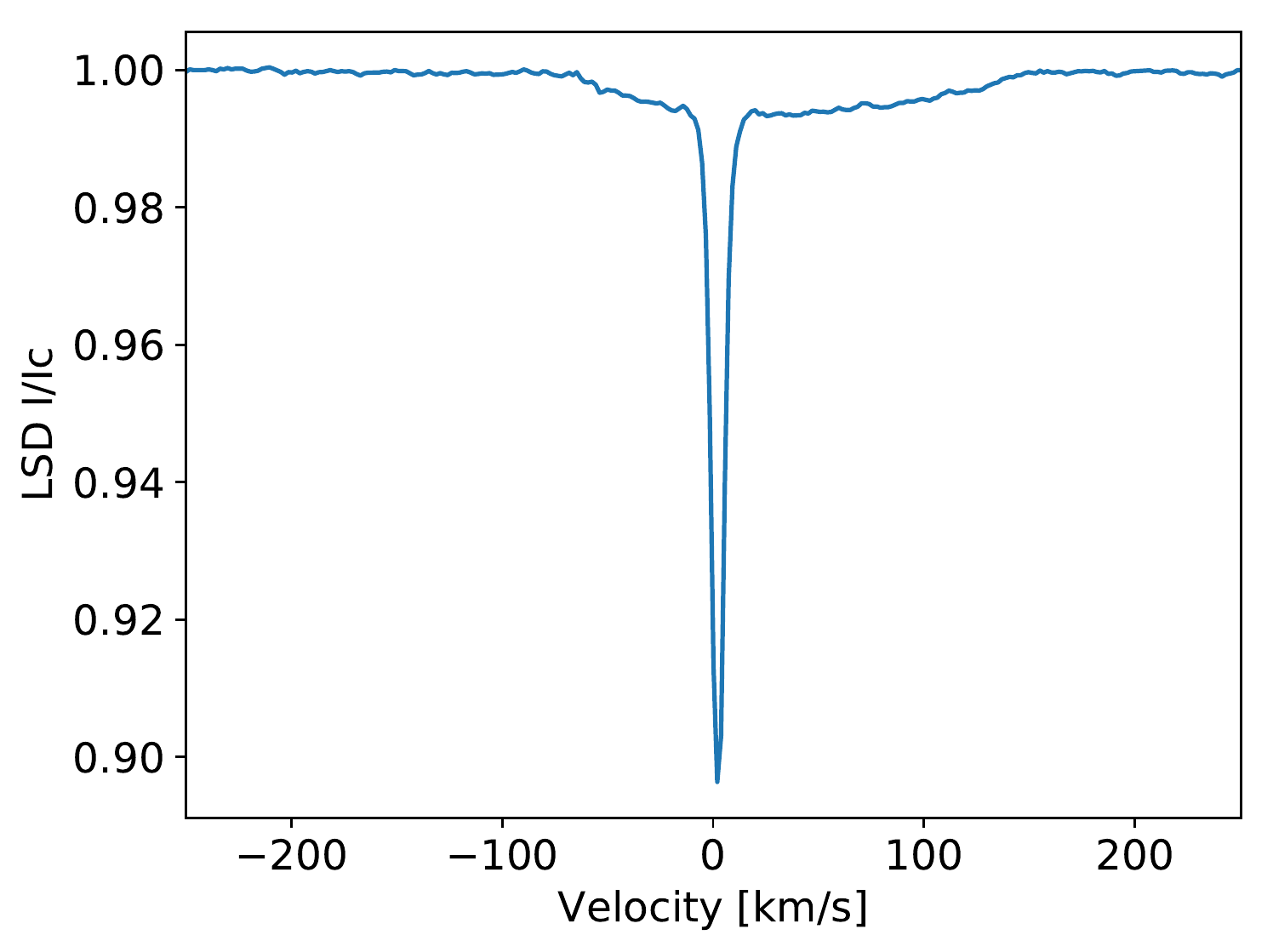}
  \caption{The LSD Stokes $I$ profile of the binary star HD\,42035.}
  \label{fig:hd42035LSD}
\end{figure}
The star HD\,42035 is a binary system. Studying the LSD Stokes $I$ 
 profile of this star (Fig.~\ref{fig:hd42035LSD}), it can be seen that the binary consists 
of a very slowly rotating component and a very rapidly rotating component. This is 
consistent with the findings of \cite{Monier2016}, who report numerous peculiarities 
of the chemical abundances, which we also observe in our spectra. We find a $T_{\rm eff}$ of 
$10500\,{\rm K}$  which is consistent with the values given in \cite{Monier2016}. However, 
as stated by \cite{Monier2016}, this value has significant flux contributions from both stars, 
and so the evolutionary status of both stars remains unclear. For this reason HD\,42035 is not shown 
in Fig.~\ref{fig:Hr}.

\subsection{HD\,186660}
For HD\,186660  we determine the $T_{\rm eff}$  to be $16900\pm300\,$K and the 
 $\log g$ to be $3.6\pm 0.2$. This suggests a star cooler than spectral type B2V \citep{Guetter1968}, and  
as can be seen from Fig.~\ref{fig:Hr} the $T_{\rm eff}$ and $\log g$ are consistent with the value of $L$ given in 
Table~\ref{Table:Lumino_Params}. As a result 
we conclude that HD\,186660 is either at the end of the MS or 
at the start of the post-MS. We calculate an age of between 33.2 and 46.5\,Myr compared to the age at the MS turn-off of between 32.9 and 52.8\,Myr, respectively. We conclude that it has a mass of 7.0--8.0\,M$_{\odot}$ and a radius of 6.5--9.4\,R$_{\odot}$. This suggests a radius increase of between a factor of 1.9 and 3.0 since the ZAMS.
\subsection{HD\,188209}
HD\,188209 is an O supergiant and so, in the same way as 15\,Sgr, our determination of $T_{\rm eff}$ and $\log g$ is focussed mainly on the analysis of {\sc uvbybeta} and the results of previous studies. We see the presence of emission in H$\upalpha$ (see Fig.~\ref{fig:alpha}), which is consistent with HD\,188209 being an O supergiant.
The values of $T_{\rm eff} = 31500^{+1000}_{-500}$\,K and  $\log g = 3.0\pm0.1$  found by \citet{Israelian2000} and $T_{\rm eff} = 29800\pm2000$\,K and  $\log g = 3.2\pm0.1$  found by \citet{Marcolino2017} agree well with our findings of $T_{\rm eff}$=$30000\pm1000\,$K and $\log g$=$3.1\pm0.2$. These values also agree well with the spectral type of O9.5Iab. 
The plots in Fig.~\ref{fig:Hr} show  
that HD\,188209 is a MS star rather than on the post-MS and we calculate an age of between 2.5 and 7.6\,Myr compared to the age at the MS turn-off of between 3.0 and 9.5\,Myr, respectively. We conclude that it has a mass of 19.7--70.0\,M$_{\odot}$ and a radius of 10.7--40.9\,R$_{\odot}$. This suggests a radius increase of between a factor of 1.9 and 3.4 since the ZAMS.
\subsection{HD\,209419}
The results by 
\citet[][$T_{\rm eff}  = 14100\,{\rm K}$ and $\log g = 3.70$]
{Lyubimkov2005} are consistent with our findings of 
$T_{\rm eff}  = 14100\pm300\,$K and  
 $\log g = 3.5\pm0.2$. These parameters suggest a star slightly cooler than the 
spectral type of B5V from \citet{Gkouvelis2016}.
The plots in Fig.~\ref{fig:Hr} show  
that HD\,209419 is either a post-MS star or at the end of the MS. We infer an age of between 110.1 and 88.1\,Myr compared to the age at the MS turn-off of between 109.2 and 88.2\,Myr, respectively. We conclude that it has a mass of 5.0\,M$_{\odot}$ and a radius of 5.2--6.4\,R$_{\odot}$. This suggests a radius increase of between a factor of 2.2 and 2.5 since the ZAMS.
\subsection{HIP\,38584}
For HIP\,38584  we determine the $T_{\rm eff}$  to be $12600\pm 300\,$K and the 
 $\log g$ to be $3.0\pm0.2$, which is consistent with the spectral type of B8II \citep{Houk1978}. 
The plots in Fig.~\ref{fig:Hr} show  
that HIP\,38584 is on the post-MS and we calculate an age of between 33.3 and 191.0\,Myr compared to the age at the MS turn-off of between 32.9 and 189.4\,Myr, respectively. We conclude that it has a mass of 4.0--8.0\,M$_{\odot}$ and a radius of 5.1--16.6\,R$_{\odot}$. This suggests a radius increase of between a factor of 2.3 and 5.2 since the ZAMS.
\subsection{HR\,3042}
Our current analysis of HR\,3042 suggests that this star is a He-weak chemically 
peculiar star. We see asymmetry in a number of 
spectral lines which could be caused by a binary, stellar pulsations, or indeed by the presence of 
a magnetic field. It is classified by \citet{Renson2009} as 
a binary because it shows variable radial velocity. However, further analysis of 
spectra from multiple rotational phases is required before we can conclude about the 
exact nature of this star. Assuming a single star, we calculate a $T_{\rm eff}$ of $14150\pm300\,{\rm K}$ and 
a $\log g$ of 3.5$\pm$0.2, which is consistent with the spectral type of B8/9II \citep{Houk1988}. The plots in Fig.~\ref{fig:Hr} show  
that HR\,3042 is most likely a post-MS star or at the end of the MS. We calculate an age of between 110.1 and 91.7\,Myr compared to the age at the MS turn-off of between 109.2 and 113.6\,Myr, respectively. We conclude that it has a mass of 5.0--5.0\,M$_{\odot}$ and a radius of 4.9--6.3\,R$_{\odot}$. This suggests a radius increase of between a factor of 1.7 and 2.5 since the ZAMS.
\subsection{PT\,Pup}
We determine $T_{\rm eff}  = 26300\pm500\,$K and $\log g = 4.1\pm0.2$ for 
PT\,Pup, which is consistent with the result of
 \citet[][$T_{\rm eff}  = 26300\pm300\,$K and $\log g = 4.15\pm0.05$]{Nieva2014} and with the spectral type B1V \citep{Nieva2013}. The plots in Fig.~\ref{fig:Hr} show  
that PT\,Pup is a MS star rather than on the post-MS and we calculate an age of between 9.1 and 6.3\,Myr compared to the age at the MS turn-off of between 18.9 and 17.7\,Myr respectively. We conclude that it has a mass of 11.0--12.0\,M$_{\odot}$ and a radius of 4.5--5.4\,R$_{\odot}$. This suggests a radius increase of between a factor of 1.2 and 1.3 since the ZAMS.
\subsection{V399\,Lac}
The $T_{\rm eff}$  ($8500\pm200\,$K) and $\log g$  ($1.5\pm0.2$) which we 
calculate for V399\,Lac match well with 
the values calculated by \citet[][$T_{\rm eff}  = 8400\pm150\,$K and $\log g = 1.40\pm0.10$]{Firnstein2012} and the spectral type A3Ib \citep{Abt1995}. 
The plots in Fig.~\ref{fig:Hr} show  
that V399\,Lac is well onto the post-MS  and we calculate an age of between 15.1 and 58.9\,Myr compared to the age at the MS turn-off of between 14.2 and 51.7\,Myr, respectively. We conclude that it has a mass of 6.9--14.4\,M$_{\odot}$ and a radius of 38.7--129.7\,R$_{\odot}$. This suggests a radius increase of between a factor of 12.5 and 24.8 since the ZAMS.
\begin{figure*}
  \includegraphics[width=.95\textwidth,height=.95\textheight]{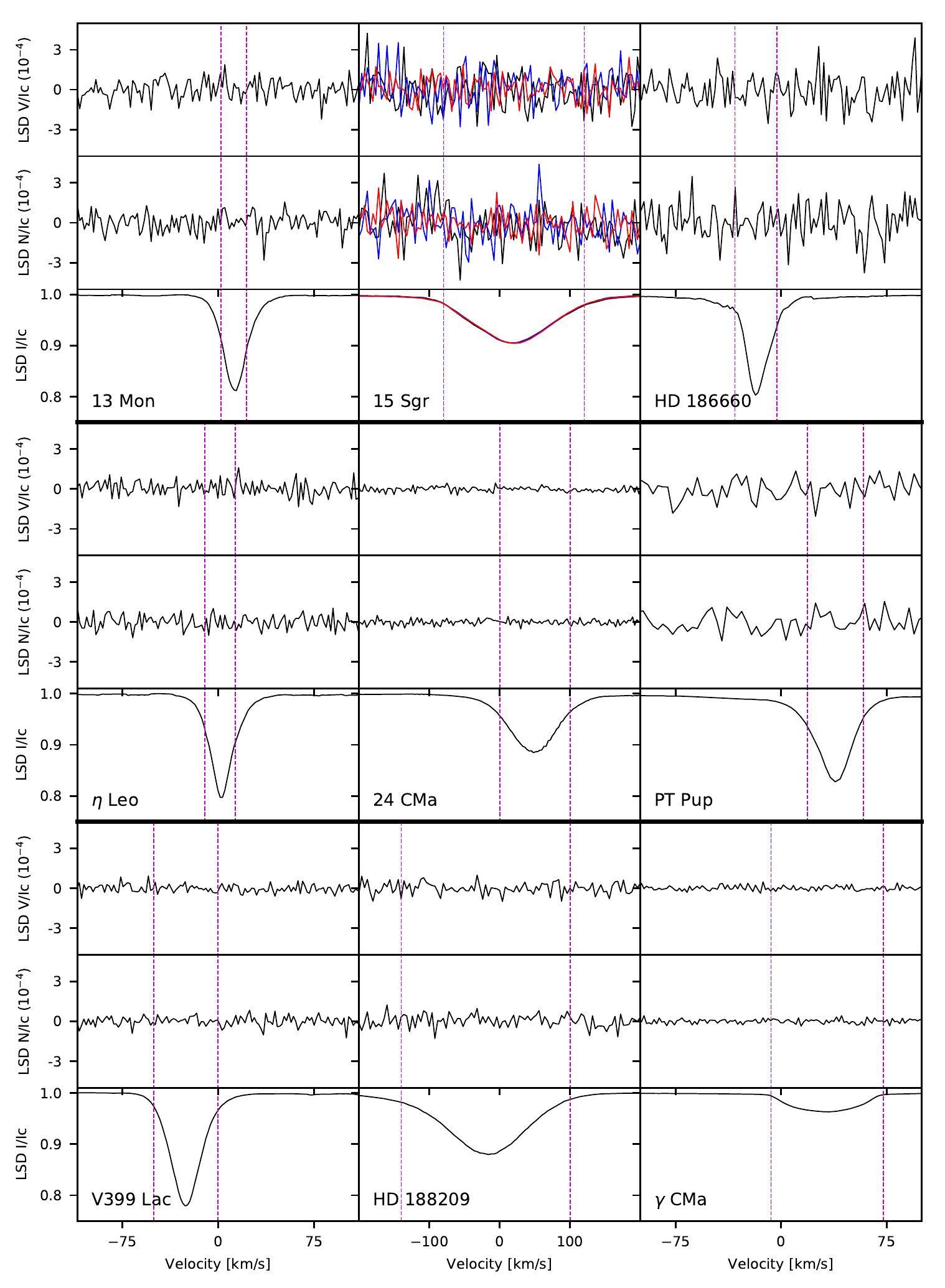}
  \caption{The LSD profiles of the LIFE targets for which no magnetic field was detected. 
                 A black solid line represents the first observation for that star in this series, 
                 blue is the second and red is the third. The dashed lines 
                 show the integration region used to calculate the magnetic field strength and FAP.}
  \label{fig:Life_Set_1}
\end{figure*}
\begin{figure*}
\ContinuedFloat
  \includegraphics[width=.95\textwidth]{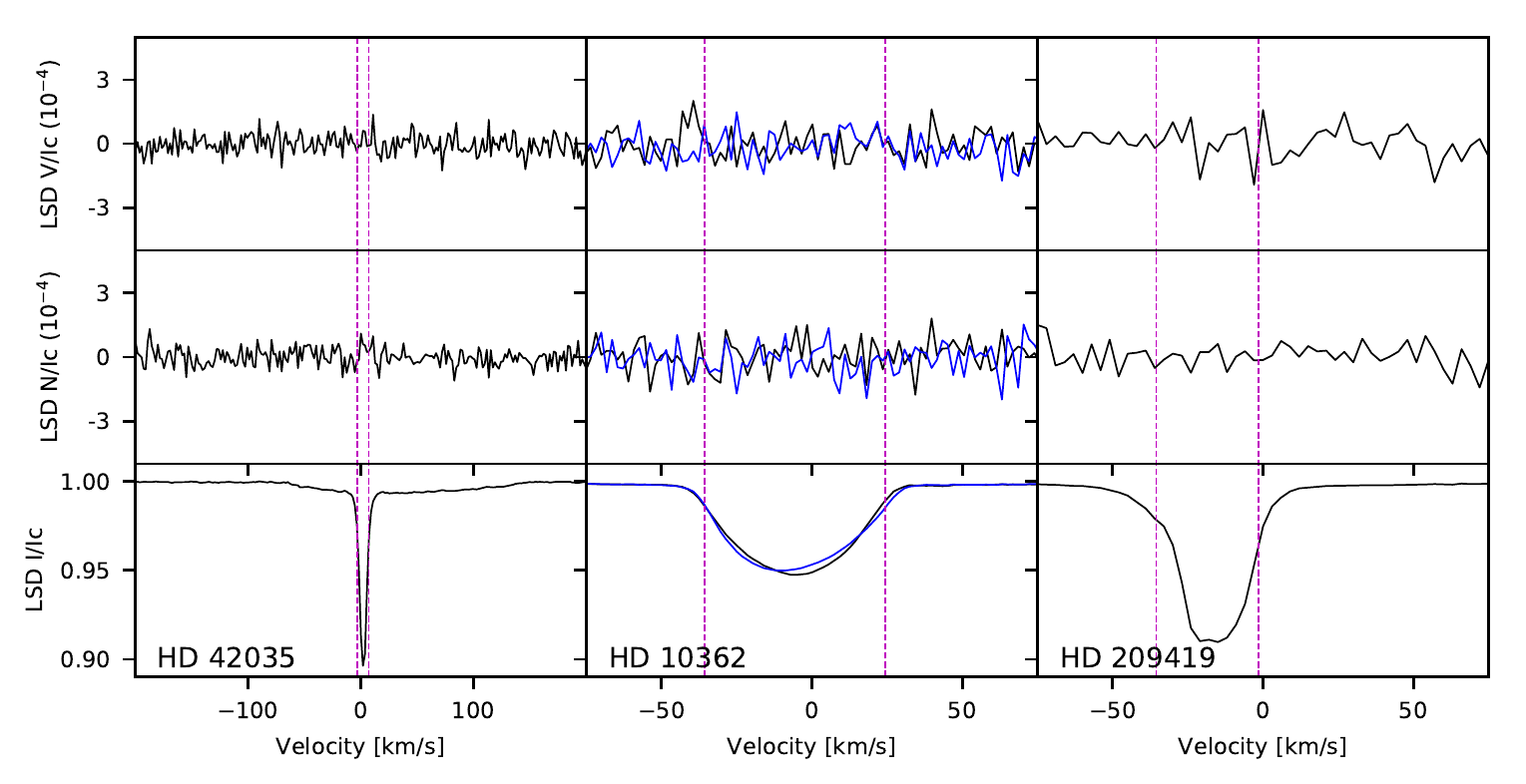}
\caption[]{{\it continued}}
  \label{fig:Life_Set_2}
\end{figure*}
\section{Magnetic Analysis and Results}
\label{sec:mag}
\begin{figure}
  \includegraphics[width=\columnwidth]{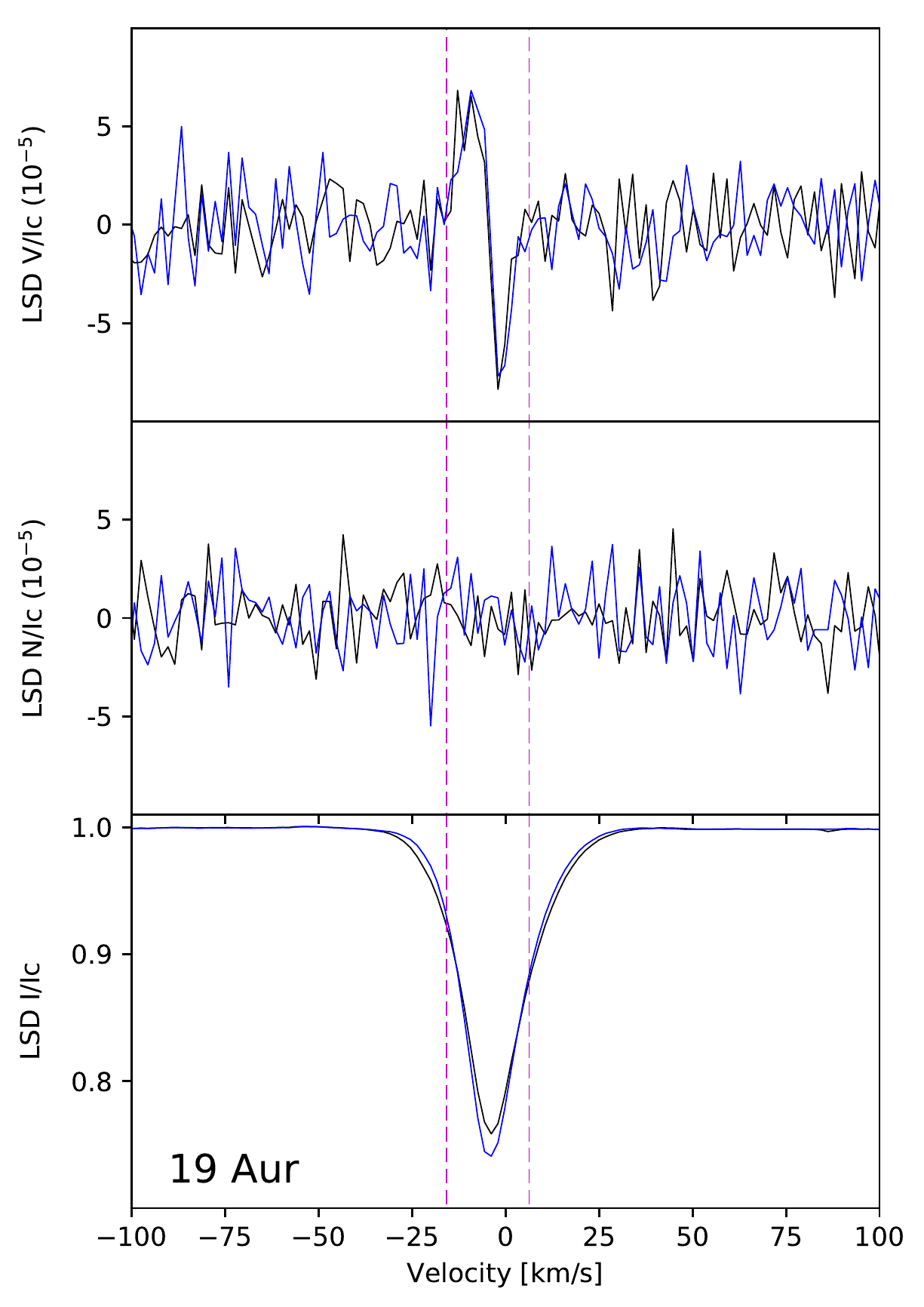}
  \caption{The LSD  Stokes $I$, $N$ and Stokes $V$ profiles of 19\,Aur, the black line is the observation 
taken on the $18^{\rm th}$ of September 2016 and the blue line is the observation taken 
on the $20{\rm th}$ of October 2016. The dashed lines 
                 show the integration region used to calculate the magnetic field strength  and FAP.}
  \label{fig:hd19aurLSD}
\end{figure}

\begin{figure}
  \includegraphics[width=\columnwidth]{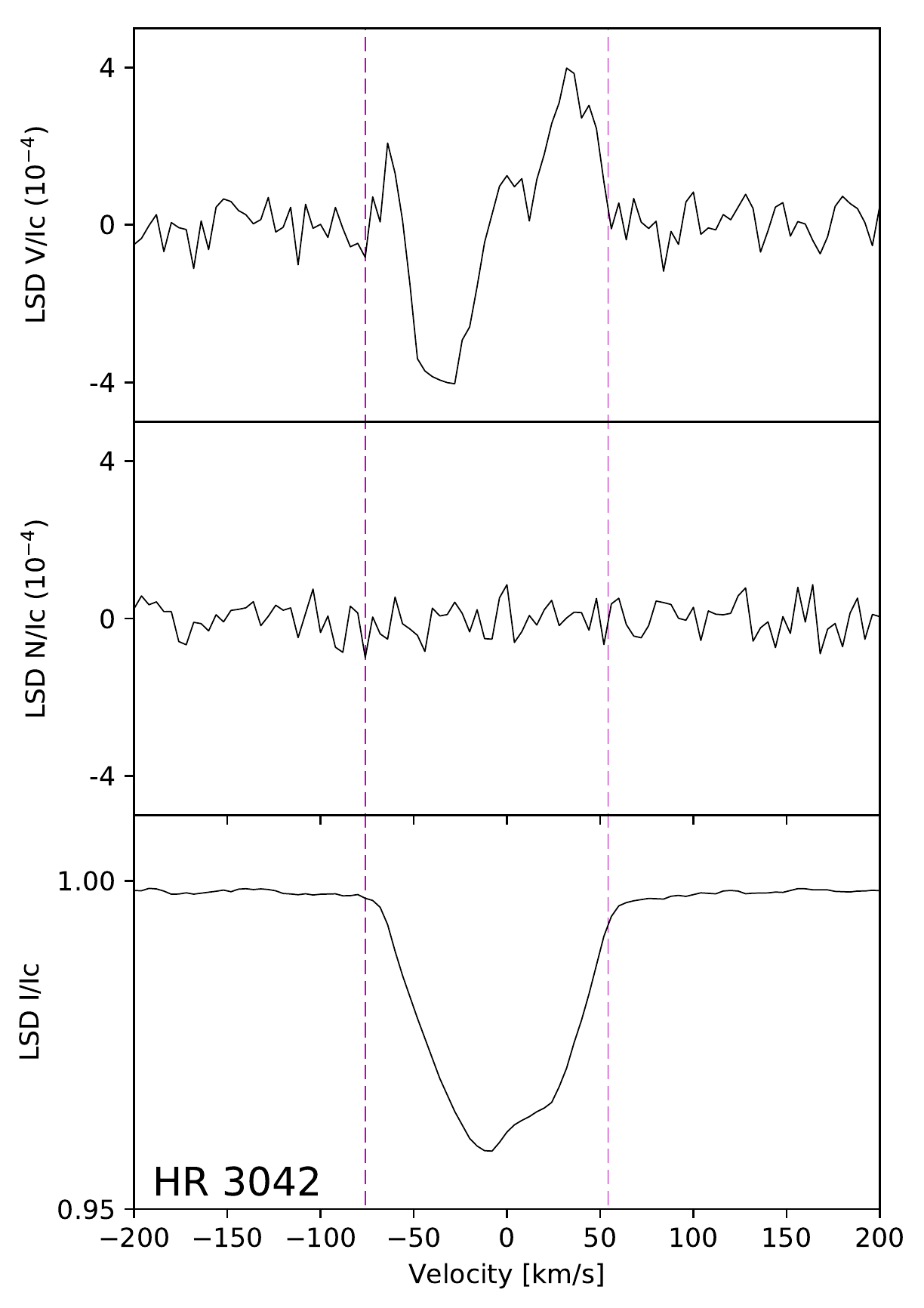}
  \caption{The LSD  Stokes $I$, $N$ and Stokes $V$ profiles of HR\,3042, observed on the 
14$^{\rm th}$ of December 2016. The dashed lines 
                 show the integration region used to calculate the magnetic field strength and FAP.}
  \label{fig:hr3042LSD}
\end{figure}

\begin{figure}
  \includegraphics[width=\columnwidth]{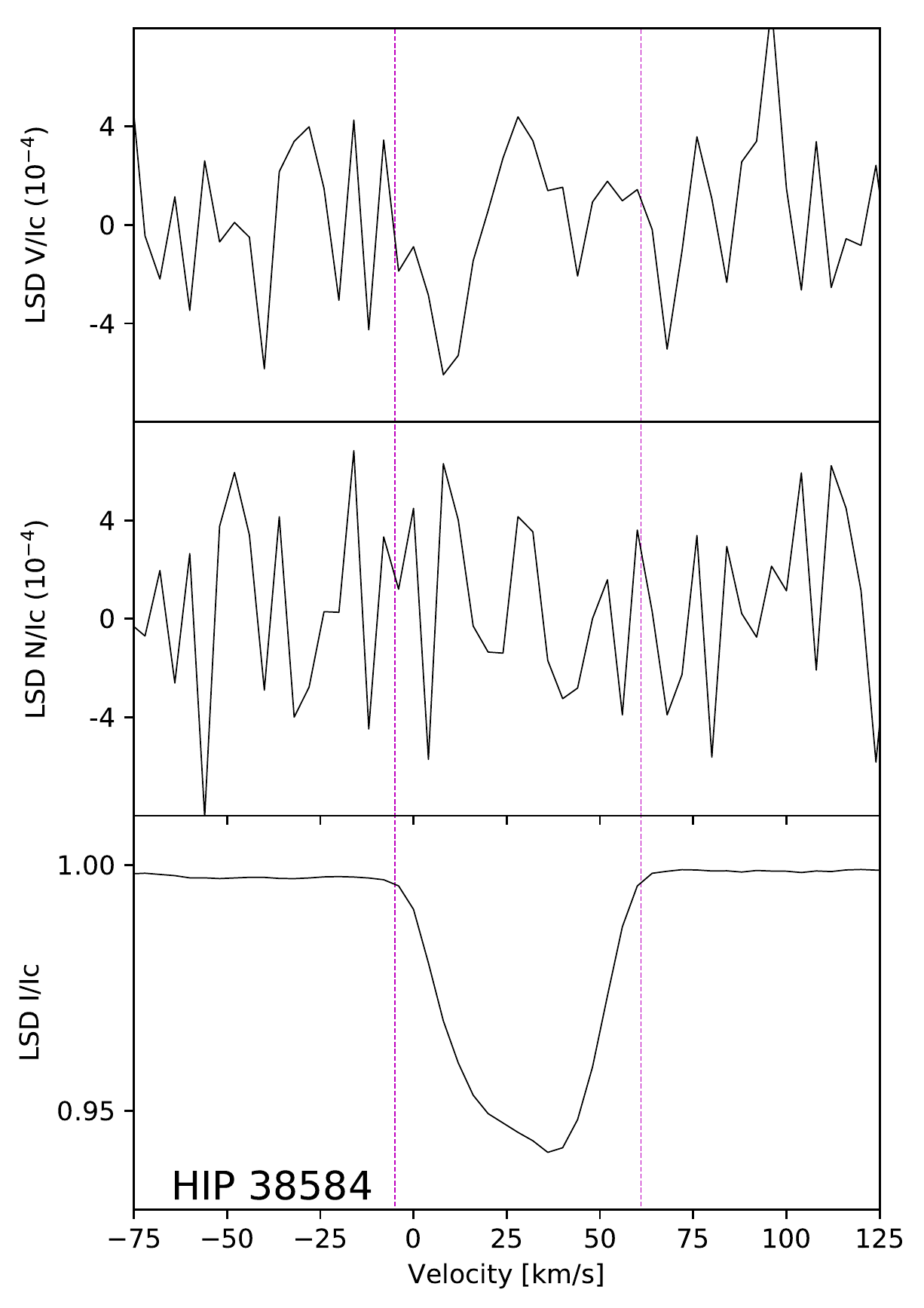}
  \caption{The LSD Stokes $I$, $N$ and Stokes $V$ profiles of HIP\,38584, observed on the 
13$^{\rm th}$ of March 2017. The dashed lines 
                 show the integration region used to calculate the magnetic field strength and FAP.}
  \label{fig:hip38584LSD}
\end{figure}

\begin{table}
	\caption{Magnetic field measurements and detection result for the observed stars. 
    The name of the star, measured longitudinal field strength ($B_l$), and 
     measured longitudinal field strength  in the null profile ($N_l$) are given. 
     Column 4 gives the results of the FAP analysis (see Section~\ref{sec:mag}) 
     where {\bf DD} is a definite field detection and ND is a null detection. 
     $B_{\rm d}$ is the current dipole field strength and $B_{\rm MS}$ is the 
     estimated dipole field strength the star would have had at the 
     ZAMS assuming only flux conservation.}        
	\label{Table:LIFE_Mag} 
	\centering                               
	\resizebox{\columnwidth}{!}{
    \begin{tabular}{@{}l@{\,\,\,}r@{\,$\pm$\,}l@{\,\,\,}r@{\,$\pm$\,}l@{\,\,\,}c@{\,\,\,}c@{\,\,\,}c@{}}       
		\hline\hline
		Star	& \multicolumn{2}{c}{$B_l$}	& \multicolumn{2}{c}{$N_l$}	& Field		& $B_{\rm d}$ (min)	& $B_{\rm ZAMS}$ (min)\\   
			& \multicolumn{2}{c}{(G)}		& \multicolumn{2}{c}{(G)}		& detect?	
& (G)	& (G)	\\
		\hline                      
             13\,Mon		&1.0	&0.6	&0.5	&0.6	& ND \\
             15\,Sgr		&0	&20	&10	&20	& ND \\
             			&0	&20	&10	&20	& ND \\
             			&20	&30	&-20	&30	& ND \\
             19\,Aur		&1.0	&0.2	&0.3	&0.2	&{\bf DD}	&     3	&   470 --   900\\
           			&1.0	&0.2	&-0.02	&0.2	&{\bf DD}	&     3	&   460 --   880\\
             24\,CMa	&2	&1	&0.4	&1.0	& ND \\
       $\upeta$\,Leo	&0.2	&0.6	&0.3	&0.6	& ND \\
     $\upgamma$\,CMa	&9	&8	&-4	&8	& ND \\
           HD\,10362	&-7	&6	&3	&6	& ND \\
				&-5	&7	&-3	&7	& ND \\
           HD\,42035	&-0.2	&0.7	&-0.1	&0.7	& ND \\
          HD\,186660	&-0.8	&2.0	&-0.4	&2.0	& ND \\
          HD\,188209	&20	&20	&20	&20	& ND \\
          HD\,209419	&5	&3	&-0.2	&3.0	& ND \\
	HIP\,38584	&3	&6	&7	&6	& ND \\
            HR\,3042	&-230	&10	&-10	&10	&{\bf DD}	&   760	&  2220 --  4810\\
             PT\,Pup	&-1	&2	&1	&2	& ND \\

           V399\,Lac	&0.9	&0.8	&0.4	&0.8	& ND \\
		\hline          	                                  
	\end{tabular}}
\end{table}
The LSD profiles calculated  for the LIFE targets (see Section~\ref{sec:LSD}) are shown in 
Figs.~\ref{fig:Life_Set_1}--\ref{fig:hip38584LSD}. For each of these profiles, we calculate 
the longitudinal magnetic field $B_l$ and the False Alarm Probability (FAP). 
We calculate $B_l$ following \cite{Rees1979} and \cite{Wade2000}: 
\begin{equation}
B_l = $-$2.14\times10^{11}\frac{\int vV(v) {\rm d}v}{\lambda zc\int \left[I_{\rm c}-I(v)\right]dv},
\end{equation}
where $v$ is the position in velocity space, $I_{\rm c}$ is the Stokes $I$ continuum level. 
$\lambda$ and $z$ are the wavelength in nm and Land\'{e} factor adopted to 
scale the LSD profiles and $c$ is the speed of light in the same unit as $v$. 
The integration limits are chosen so that they do not extend beyond the Stokes $I$ line profile and so that 
they avoid the wings of the supergiants. This is because in supergiants the line wings may be formed in the stellar wind and so are not sensitive to the surface field. This is consistent with the Stokes $V$ signatures presented by \citet{Neiner2017}.

The FAP is a quantity used to determine the probability that a Stokes $V$ signature is real or 
noise. It is calculated using the $\upchi^2$ probability function to ascertain 
whether the deviation from zero observed in the Stokes $V$ and $N$ profiles is best explained as a result of random 
noise or a signal \citep{Donati1992,Donati1997}. 
We follow the convention of \citet{Donati1997}, who 
define  a definite detection to be a probability (P) that the signal is real of 
at least 99.999\% which is a FAP of $10^{-5}$, a marginal detection is a 
99.999\% \textgreater\ P \textgreater\ 99.9\%  ($10^{-5}$ \textless\ FAP 
\textless\ $10^{-3}$) and no detection is P \textless\ 99.9\% (FAP 
\textgreater\ $10^{-3}$). 
We require that 
the signal is only detected in Stokes $V$ not in $N$, and that the 
Stokes $V$ signal is contained within the width of the LSD Stokes $I$ profile. 

The $B_l$ and FAP values are listed in Table~\ref{Table:LIFE_Mag}. Comparing these results with the expected field strengths shown in Table~\ref{Table:Lumino_Params}, we see that we reach a magnetic precision sufficient to detect the expected field in all cases.
Our results show 
the clear detection of a magnetic field in two of the stars: 
19\,Aur and HR\,3042. This is also clearly observed in their LSD profiles shown in Figs.~\ref{fig:hd19aurLSD} and \ref{fig:hr3042LSD}. Moreover, HIP\,38584 shows a possible Zeeman signature (see Fig.~\ref{fig:hip38584LSD}) which is consistent with the velocity range of the intensity line. However, its FAP leads to a non-detection and its $B_l$ value is compatible with 0. The observations of all the other stars result in non-detections and do not show any signs of coherent structures in their Stokes $V$ profiles. Below we discuss in more detail the 2 magnetic stars and the magnetic candidate.
\subsection{The magnetic star 19\,Aur}
We observed 19\,Aur twice, on the 18$^{\rm th}$ of September and 20$^{\rm th}$ of October 2016. 
Our observations consisted of five consecutive Stokes $V$ sequences of four 
subexposures each with an exposure time of 254 seconds, resulting in a total 
exposure time of 1.41 hours for each combined profile. For the line mask
 4298 lines were included after rejections based on the constraints described in Section~\ref{sec:LSD}.

We calculate $B_l$ over an integration range of $\pm 10$\,kms$^{-1}$ about the line center of 
$-3.1$\,kms$^{-1}$ for the Stokes $I$ and $V$ and $N$ profiles. This results in a definite field detection 
for each observation sequence, 
both with a longitudinal field of $1.0\pm0.2$\,G measured from the Stokes $V$ profile and no detection in 
 the $N$ profile. The LSD profile of 19\,Aur shows a clear antisymmetric Stokes $V$ signature contained within 
 the line center and an essentially flat null profile, providing further evidence for a 
magnetic field in this star.

The two Stokes V profiles of 19\,Aur are very similar in shape, which suggests that the rotational period is 
either very long (much longer than the one month that elapsed between the two observations) or by chance we observed the star in close to the same 
phase in both observations (suggesting a rotational period of $\sim$ 33 days  or a sub-multiple of 33 days). A third possibility is that either $i$ and/or $\beta$ are close to zero resulting in a invariable Stokes $V$ signature. Based on the current 
radius and $v\sin i$ of 19\,Aur we estimate its rotational period to be 210\,--340\,days and thus we favour the first explanation.

We estimate $B_{\rm d} > 3$\,G,  which, given a current radius of between 36.8 and 59.3 $R_\odot$, suggests an 
estimated $B_{\rm ZAMS}$ of between 460 and 900\,G. These values are entirely consistent with the range of magnetic field 
values observed in magnetic MS B stars.

\subsection{The magnetic star HR\,3042}
We observed HR\,3042 once on the 14$^{\rm th}$ of December 2016.
Our observations consisted of three consecutive Stokes $V$ sequences of four 
subexposures each with an exposure time of 981 seconds, resulting in a total 
exposure time of 3.27 hours for the combined profile. For the line mask
 873 lines were included after rejections based on the constraints described in Section~\ref{sec:LSD}.

We calculate $B_l$ over an integration range of $\pm 65$\,kms$^{-1}$ about the line center of 
$-5$\,kms$^{-1}$ for the Stokes $I$ and $V$ and $N$ profiles. This  results in a definite field detection, 
with a longitudinal field of $-230\pm10$\,G measured from the Stokes $V$ profile and no detection in 
 the $N$ profile. The LSD profile of HR\,3042 shows a very strong  Stokes $V$ signature 
and a flat null profile. This adds further evidence for a 
magnetic field in this star.

The Stokes $I$ line profile shows an asymmetry at the core, this could be an effect of 
the magnetic field, stellar pulsations, binary companion or the presence of surface spots. Further observation will 
allow us to determine the nature of this magnetic star.

We estimate $B_{\rm d} > 760$\,G, which, given a current radius of between 4.9 and 6.3\,$R_\odot$, suggests an 
estimated $B_{\rm ZAMS}$ of between 2220 and 4810\,G. Thus, HR\,3042 was likely a quite strongly magnetic star at the start of the MS, and is or was possibly an Ap/Bp star. 
Based on the current 
radius and $v\sin i$ of HR\,3042 we estimate its rotational period to be 4.1\,--5.3\,days. 
\subsection{HIP\,38584: a magnetic candidate}
We have observed HIP\,38584 once,  on the 13$^{\rm th}$ March 2017. 
Our observations consisted of one Stokes $V$ sequence of four 
subexposures each with an exposure time of 1049  seconds, resulting in a total 
exposure time of 1.17 hours for the combined profile. For the line mask
 864 lines were included after rejections based on the constraints in Section~\ref{sec:LSD}.

The FAP analysis results in no detection of a magnetic field. However, the line profile 
of HIP\,38584 is asymmetric which could suggest a binary companion or the presence of surface spots. There is evidence of 
a coherent structure in half of the line profile, which could suggest that one of the 
two stars is magnetic. Further observations of this star are therefore necessary to 
determine whether or not a magnetic field is present. If this star is indeed a binary and magnetic, it will provide valuable 
 data for the BinaMIcS \citep{Alecian2015} project, which seeks to understand the magnetism of close binary  stars.
 \section{Discussion}
\label{sec:dis}

Two of the stars we have observed show a clear Stokes $V$ signature and a third shows a possible 
Zeeman signature in Stokes $V$ and requires further observations to check its potential 
magnetic nature. Of the stars observed, six are almost certainly post-MS, five are either 
at the end of the MS or at the start of the post-MS and three are MS stars. The evolutionary status of HD\,42035 remains unclear. 
This leads to an incidence rate of magnetic fields in post-MS stars of between 10 and 29\%, which is compatible with the $\sim$10\% incidence on the MS. 
However, our 
sample  of post-MS stars is so far insufficient for these values to have any statistical significance. The full LIFE sample will be necessary to draw clear conclusions.

Of the non-magnetic stars, 15\,Sgr was previously studied as part of the MiMeS survey \citep{Grunhut2017}. The authors obtained only one definite field detection, which upon a visual inspection of the 
LSD profiles was shown to result from a coherent structure extending outside of the line profile. As a consequence, it 
was determined to be spurious. Our 
observations consist of two sets of 2$\times$4$\times$344s exposures and one of 
4$\times$4$\times$344s compared with 3$\times$4$\times$300s \citep{Grunhut2017}. This gives us a 
higher overall S/N, and so the potential to detect weaker fields. Our additional 
observations show no evidence of a magnetic field detection or the aforementioned signature, 
confirming that it was almost certainly spurious. In addition we do not detect a mean longitudinal magnetic 
field for $\upgamma$\,CMa, confirming the result of \citet{Hubrig2012}. 

Of the two clearly magnetic stars, one is very close to or just past the 
turn-off point from the MS to the post-MS (HR\,3042) and one is a clearly evolved 
star (19\,Aur). HR\,3042 has a current mass of 5\,$M_\odot$, a $T_{\rm eff}$ of 
14000$\pm$300\,K and shows an underabundance of He. Its current $B_{\rm d} > 760$\,G, in combination with a predicted radius 
expansion of a factor between 1.8 and 2.6 suggests a magnetic field strength of at least 
2220 and 4810\,G at the ZAMS if magnetic flux conservation is the only process which 
affects the evolution of the magnetic field strength.

19\,Aur has a current mass of 6.9-9.7\,$M_\odot$ and a $T_{\rm eff}$ of 
8500$\pm$200\,K and we see approximately solar 
abundances of elements consistent with the results of \citet{Lyubimkov2015}. Its current $B_{\rm d} > 3$\,G, in combination with a predicted radius 
expansion of a factor between 12.0 and 16.5 suggests a magnetic field strength of at least 
460--900\,G at the ZAMS if magnetic flux conservation is the only process which 
affects the evolution of the magnetic field strength. With our observations so far we are unable 
to determine the type of field which exists in this star. However, since we have two observations taken 33 days apart that show essentially the same signature, there is good evidence that this star hosts only a fossil field.

The detailed magnetic and chemical abundance analysis of both HR\,3042 and 19\,Aur, using 
spectra at multiple rotational phases, will  
allow us to better constrain the 
mass, radius, metallicity,  age and field geometry for each of these stars.

Combining our findings with those of \citet{Neiner2017}, there are now three known clearly evolved magnetic  A-type stars: 19\,Aur, $\upiota$\,Car and HR\,3890.
The $B_{\rm ZAMS}$ calculated for these stars using flux conservation is   
consistent with distribution of magnetic field strengths on the MS for hot stars 
\citep[][and Shultz et al., in prep.]{Shultz2016}. However, these results do not exclude other possible magnetic field decay mechanisms.

\section{Conclusions}
\label{sec:conc}
In this paper, we have presented the first observations and results of the LIFE project. 
Out of our sample of 15 stars (six post-MS, five MS/post-MS and three MS stars)\footnote{The evolutionary status of HD\,42035 remains unclear because it is a binary with significant flux contributions from both stars.}, we found two to be magnetic HR\,3042 and 19\,Aur and a magnetic candidate, HIP\,38584. 
 HR\,3042 is either at the very end of its MS lifetime or 
just onto the post-MS and 19\,Aur is very definitely post-MS. We find that the less evolved 
star, HR\,3042, has $B_l = -230\pm10$\,G and the more evolved 19\,Aur 
has $B_l = -1.0\pm0.2$\,G. This is consistent with what would be expected if the field strength 
has decreased only as a result of flux conservation. It is important to note that the work by \citet{Fossati2016} found that magnetic flux decay likely has a significant impact on the evolution of magnetic fields. Therefore the estimated ZAMS magnetic field strengths we calculate in this paper may be underestimated. Further analysis of the magnetic fields in evolved OBA will allow us to further understand the impact of possible magnetic flux decay.

Studying 19\,Aur and HR\,3042 in detail, using spectropolarimetric observations 
at multiple phases in each star's rotation period, will allow us to determine whether 
these stars still host fossil fields or  whether a dynamo field has 
formed in the outer convective region which is modifying the fossil field. This would 
likely be evidenced by changes in the line profiles for the same rotation phase in different 
epochs. In addition, the continuation of the LIFE project is predicted to yield at least six 
further magnetic   post-MS stars, if the prevalence of magnetic fields in these stars 
is consistent with their MS counterparts.

The results of the LIFE project will provide new insight on the nature of the interplay between magnetic fields and stellar evolution. The results will also provide additional evidence for theories which aim to connect the magnetic fields observed in MS stars with those observed in the later stages such as white dwarfs, neutron stars and black holes.
\\

\section*{Acknowledgements}

SM acknowledges support from ERC SPIRE grant (647383).  
WM acknowledges support from CNPq, grant 307152/2016-2. 
GAW acknowledged support from the Natural Sciences and Engineering Research Council (NSERC) of Canada in the form of a Discovery Grant. 
This research has made use of the SIMBAD database operated at CDS, Strasbourg 
(France), and NASA's Astrophysics Data System (ADS).
This work has made use of data from the European Space Agency (ESA)
mission {\it Gaia} (\url{https://www.cosmos.esa.int/gaia}), processed by
the {\it Gaia} Data Processing and Analysis Consortium (DPAC,
\url{https://www.cosmos.esa.int/web/gaia/dpac/consortium}). Funding
for the DPAC has been provided by national institutions, in particular
the institutions participating in the {\it Gaia} Multilateral Agreement.


\bibliographystyle{mnras}
\bibliography{LIFE.bib} 
\begin{figure*}
  \includegraphics[width=\textwidth]{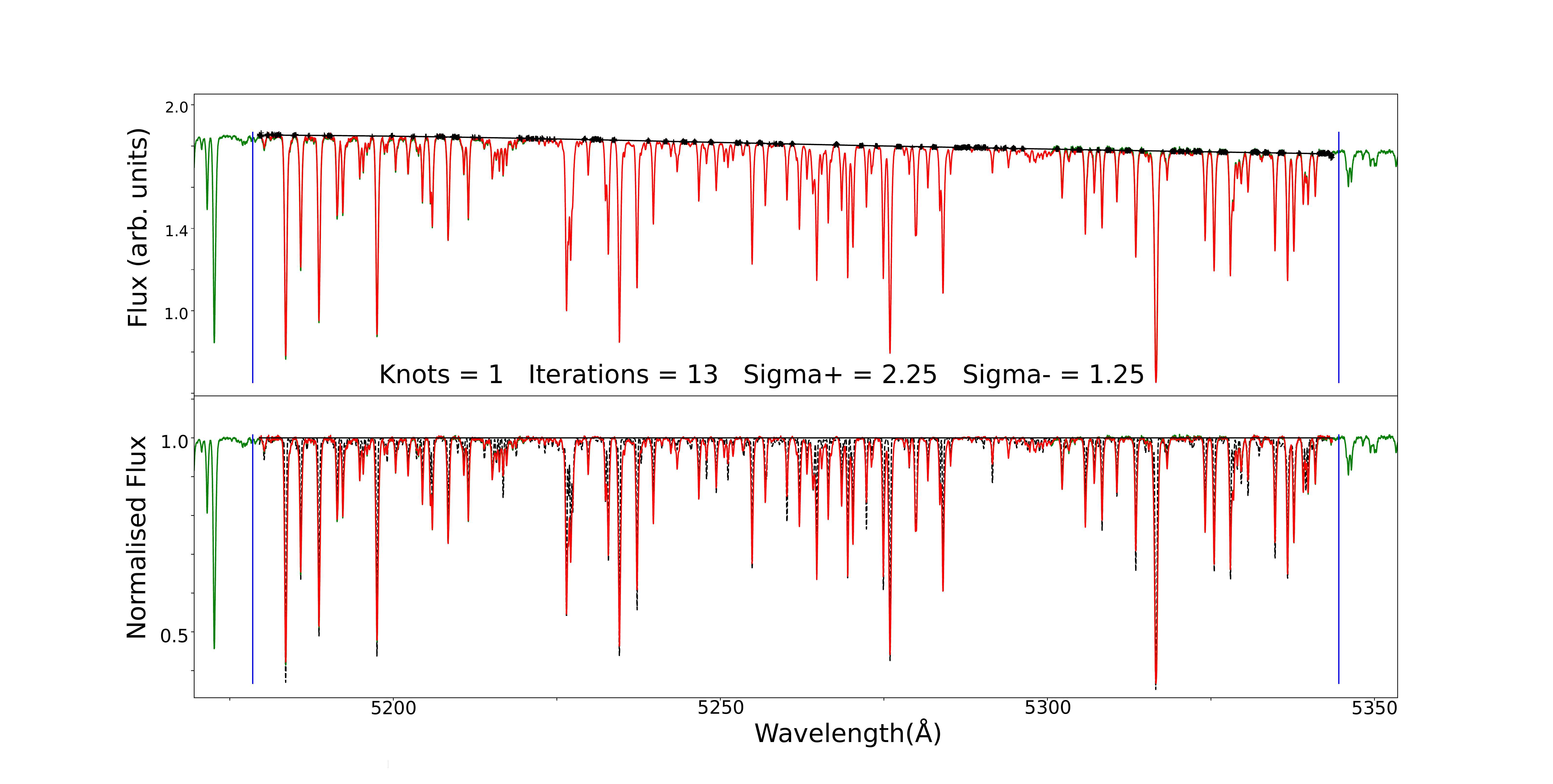}
  \caption{The graphical interface for {\sc spent}. The upper and lower panel of the interface shows one order of the 
unnormalised and normalised echelle spectrum respectively (red solid line), with a portion of the preceding and 
succeeding orders (green solid lines). The black pluses in the upper panel are the points calculated to be at 
continuum level based on the current input parameters. The solid black line is the cubic 
spline fit to these continuum points in the upper panel and a line at unity in the lower panel. The blue lines show the edge regions where points can be removed and replaced by a cluster of points. The current values for the input parameters are also shown and finally the dashed black line is a synthetic spectrum 
calculated with the fundamental parameters associated with the current star.}
  \label{fig:hrNormaliser}
\end{figure*}
\appendix

\section{Normalisation Code}
\label{appendix:normal}
The LIFE project 
involves the analysis of a large number of spectra. Part of this analysis involves 
fitting the continuum of each spectral order for each observation to scale the 
stellar continuum to unity. Generally this is 
accomplished using {\sc IRAF}, which is a very powerful tool. However, since we 
have such a large number of spectra our aim is to accomplish the normalisation 
in a semi-automatic way. As a result, we have developed a new code, 
{\sc spent} (SPEctral Normalisation Tool), a {\sc python} code which
combines automatic sigma-clipping with adjustable edge regions in an 
interactive interface. An example of the graphical user interface of {\sc spent} 
is shown in Fig. \ref{fig:hrNormaliser}. 

The code begins by fitting a third-order spline to one order of an 
unnormalised echelle spectrum. The points which make up the spectrum are 
then iteratively $\upsigma$-clipped until the third order spline only passes 
through continuum points and so models the continuum. In this paper we only have 
 absorption spectra, and so we must clip more points below the fit than 
above, to remove the absorption lines. However, the reverse is equally possible. The user is interactively able to define the 
number of interior knots used to calculate the spline fit, the number of iterations of the 
$\upsigma$-clipping and the upper and lower $\upsigma$ threshold values.
Each time a change is made to the parameters, the normalised spectrum is 
updated. 
For each star the normalisation parameters are saved in a log file, this means it is possible 
to fit future observations efficiently and consistently.

A key consideration when designing this code was that 
the edges of the orders of echelle spectrograph can occur in the middle of 
spectral lines. As a result it can be difficult to normalise these regions properly. 
To help overcome this issue, {\sc spent} is built with two solutions. Firstly, the user is 
able to remove a number of points  at either or both ends of the unnormalised spectral order. A cluster of 
points at the beginning and/or end of each order replaces these. This cluster can then be 
move up and down which artificially adjusts the end knots of the spline. The second solution is to show a portion of the spectral orders at each side of the currently 
active order. This makes it possible for the user to see the characteristics of the spectral line which has been split by the orders and compensate appropriately.

\bsp	
\label{lastpage}

\end{document}